\documentclass{sn-jnl}
\usepackage[sort&compress,nonamebreak,sectionbib,square,super]{natbib}
\usepackage{setspace} 

\usepackage{graphicx}%
\usepackage{multirow}%
\usepackage{amsmath,amssymb,amsfonts}%
\usepackage{amsthm}%
\usepackage{mathrsfs}%
\usepackage[title]{appendix}%
\usepackage{xcolor}%
\usepackage{textcomp}%
\usepackage{manyfoot}%
\usepackage{booktabs}%
\usepackage{listings}%
\usepackage{titlesec}
\usepackage{xr}
\usepackage{xr-hyper}
\usepackage{hyperref}
\usepackage{lineno}

\usepackage{titletoc}

\hypersetup{
	hidelinks,
	colorlinks=true,
	linkcolor=blue,
	filecolor=blue,
	citecolor=blue,
	urlcolor=blue,
	breaklinks=false
}

\makeatletter
\newcommand*{\addFileDependency}[1]{
	\typeout{(#1)}
	\@addtofilelist{#1}
	\IfFileExists{#1}{\typeout{File #1 O.K.}}{\typeout{No file #1.}}
}
\makeatother

\begin{document}
\title[Article Title]{Coherence Awareness in Diffractive Neural Networks}

\author[1]{\fnm{Matan} \sur{Kleiner}}\email{matankleiner@campus.technion.ac.il}
\author[2]{\fnm{Lior} \sur{Michaeli}}\email{liormic1@caltech.edu}
\author*[1]{\fnm{Tomer} \sur{Michaeli}}\email{tomer.m@ee.technion.ac.il}

\affil[1]{\orgdiv{Faculty of Electrical and Computer Engineering}, \orgname{Technion}, \orgaddress{\city{Haifa}, \postcode{32000}, \country{Israel}}}

\affil[2]{\orgdiv{Division of Engineering and Applied Science}, \orgname{California Institute of Technology}, \orgaddress{\street{1200 E. California Avenue}, \city{Pasadena}, \postcode{91125}, \state{CA}, \country{USA}}}

\abstract{
Diffractive neural networks hold great promise for applications requiring intensive computational processing. Considerable attention has focused on diffractive networks for either spatially coherent or spatially incoherent illumination. 
Here we illustrate that, as opposed to imaging systems, in diffractive networks the degree of spatial coherence has a dramatic effect. In particular, we show that when the spatial coherence length on the object is comparable to the minimal feature size preserved by the optical system, neither the incoherent nor the coherent extremes serve as acceptable approximations. 
Importantly, this situation is inherent to many settings involving active illumination, including reflected light microscopy, autonomous vehicles and smartphones. Following this observation, we propose a general framework for training diffractive networks for any specified degree of spatial and temporal coherence, supporting all types of linear and nonlinear layers. Using our method, we numerically optimize networks for image classification, and thoroughly investigate their performance dependence on the illumination coherence properties. 
We further introduce the concept of coherence-blind networks, which have enhanced resilience to changes in illumination conditions. Our findings serve as a steppingstone toward adopting all-optical neural networks in real-world applications, leveraging nothing but natural light.
}

\maketitle

\doublespacing

\section{Introduction}
\label{introduction}
Optical computing has long been considered a potential alternative to electrical computing thanks to its speed, parallel nature, and energy efficiency \citep{caulfield2010future, athale2016optical, mcmahon2023physics}. These advantages were recognized as particularly important for the compute-intensive inference stage of deep neural networks, and have thus drawn much attention over the last few years \citep{shen2017deep, tait2017neuromorphic, Lin2018, chang2018hybrid, hamerly2019large, yao2019intelligent, feldmann2019all, zhang2019artificial, qian2020performing, wetzstein2020inference, shastri2021photonics, wang2022optical, fu2023photonic, cheng2024photonic}. 
In the specific context of visual data, a promising direction is that of all-optical diffractive neural networks \citep{Lin2018}, which are optical systems composed of diffractive layers, usually in the forms of phase masks. Each layer in a diffractive network comprises a grid of pixels that play the roles of neurons in digital neural networks, the phases of which are determined through an optimization (training) process to achieve a given objective. 

Diffractive networks have proven a powerful tool for diverse machine learning tasks, from image classification to imaging through scattering media \citep{zhou2021large, li2021real, kulce2021all, luo2022computational, luo2022cascadable, mengu2022classification, bai2022image, chen2023photonic}. As these networks rely solely on passive phase manipulations, they can preserve the incoming optical energy throughout the system and do not require any external energy source. However, diffractive networks have predominantly been designed to function with monochromatic spatially coherent illumination, from terahertz frequencies\citep{Lin2018, luo2022computational, bai2022image}, to near-infrared \citep{goi2022direct, bai2023data} and visible light\citep{yan2019fourier, chen2021diffractive, chen2023photonic}. Such illumination conditions, inherent to lasers, are different from those encountered in every-day settings. Indeed, the spatial and temporal degrees of coherence of \textit{e.g.} sunlight, LEDs and fluorescent lamps are always finite. 
Several works addressed the other extreme of incoherent light. In particular, in the context of spatial coherence, Rahman et al.~\citep{rahman2023universal} studied linear diffractive networks for fully incoherent illumination. However, in practice, the spatial coherence length of the illumination is never zero either. For example, the spatial coherence length of direct sunlight was measured to be $80 \lambda$, which corresponds to 32 $\mu$m -- 56 $\mu$m for visible light in the wavelength range of 400 nm -- 700 nm \citep{divitt2015spatial}. 

It is well known that in imaging systems, the degree of spatial coherence has a limited effect on the image quality, as it usually only affects the level of blurriness or halos around object boundaries
\citep{thompson1969iv, goodman2015statistical, deng2017coherence}. 
However, the optical transfer functions implemented by diffractive networks are typically very different from those of imaging systems.
Therefore, the common wisdom concerning the role of spatial coherence in imaging systems is not readily transferable to diffractive neural networks.

Here, we show that in sharp contrast to imaging systems, the degree of spatial coherence often has a dramatic effect on diffractive networks.
Specifically, when an optical system comprises elements with abrupt phase changes, as is often the case in diffractive networks, the intensity pattern at the output plane changes rapidly with the degree of coherence at the object plane. In such cases, there may be a broad range of coherence levels under which neither the fully coherent nor the fully incoherent approximations hold.
In particular, we show that in many real-world applications that involve an incoherent source (\textit{e.g.} LED), the spatial coherence length on the object is such that the system cannot be accurately modeled as operating under fully incoherent illumination. These settings arise when the optical resolution of the system is comparable to the spatial coherence length on the object, and are inherent to \textit{e.g.}~autonomous vehicles and to smartphones, where the illumination source is located near the camera's entrance pupil and is of similar dimensions.
It is important to note that the spatial coherence on the object plane cannot be controlled in such settings. Therefore, the design of diffractive networks for these applications necessitates taking the precise coherence level into account. 

To account for this effect, here we propose a coherence-aware framework, which allows training diffractive networks for illumination of any prescribed spatial and temporal degrees of coherence. Specifically, following the van Cittert–Zernike theorem, we control the degree of spatial coherence by tuning the size and distance from the object of a spatially incoherent source. To train the network, we separately propagate the electric field from each point on the source through the entire network, while properly taking into consideration intensity-dependent nonlinear layers. 

Using our new framework, we numerically investigate the optimal performance of networks for classification tasks under varying degrees of spatial and temporal coherence. We examine both the nonblind case, in which the coherence parameters are precisely known at train time, and  coherence-blind networks, which are oblivious to the precise levels of coherence. The nonblind configuration serves as an upper bound on the performance expected in practical settings, while the blind configuration can be used to deduce a lower bound by taking the range of coherence levels at train time to contain that foreseen at inference time. Interestingly, we observe that the performance in both settings improves as the light becomes more spatially coherent. 
Not surprisingly, nonblind networks achieve superior performance to their blind counterparts when the coherence conditions at inference time match those at train time. However, their performance rapidly deteriorates as the coherence at inference time deviates from the coherence for which they were trained. 
Interestingly, we observe that a mismatch in the spatial coherence can lead to a substantial drop in performance, while diffractive networks tend to exhibit greater tolerance to mismatches in the level of temporal coherence.

\section{Results}
\label{results}
\subsection{The effect of spatial coherence on diffractive networks}
\label{sec:effect_on_diff}

Before delving into our coherence-aware training framework, we start by illustrating the effect of the spatial coherence of the illumination on the output of an optical system. We particularly contrast the case where the system performs imaging with the case where the system is similar in nature to a diffractive network. Here, we use the term \textit{coherence length} to indicate the square root of the spatial coherence area~\cite{goodman2015statistical} (see precise definition in Section~\ref{sec:framework}). This is in contrast to the frequent use of this term to indicate temporal coherence (where it is taken to mean coherence time multiplied by the speed of light).

\begin{figure}
  \centering
   \includegraphics[width=0.9\linewidth]{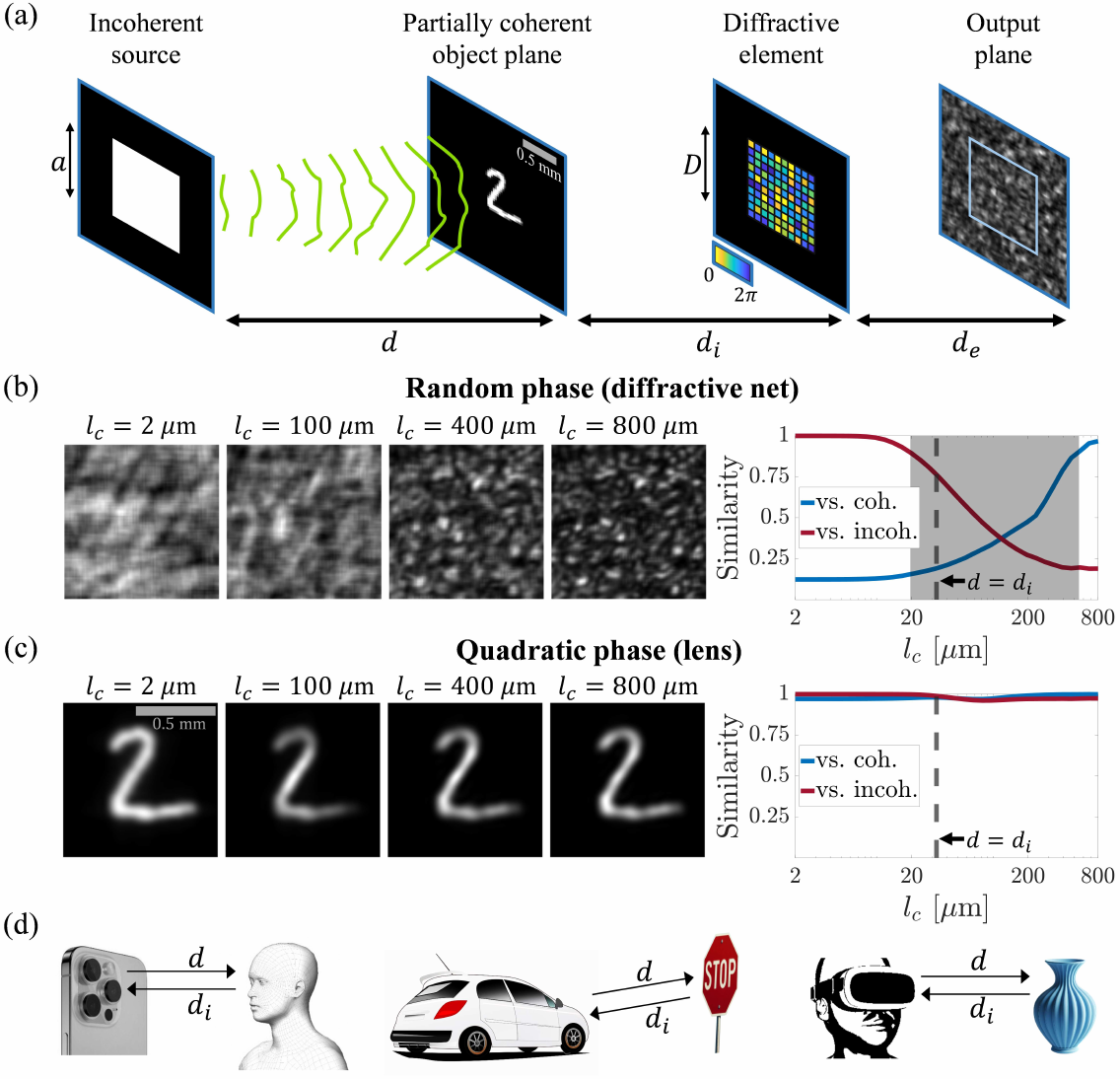}
   \caption{\textbf{The influence of the degree of spatial coherence on the output of non-imaging optical systems}. (a) A non-imaging optical system. 
   (b) The left pane depicts intensity patterns resulting from the optical system illustrated in (a), for a diffractive element whose pixels are drawn independently at random in the range $[0,2\pi]$. The object is a digit of dimensions 0.84 mm $\times$ 0.84 mm. The intensity patterns are calculated for 35 different values of $d$ which translate to different coherence lengths between 800 $\mu$m and 2 $\mu$m. All the results are for the case where $a=D$ and $d_i$ is the same across experiments. The plot on the right shows the centered cosine similarity between the intensity pattern at the output plane and the intensity pattern corresponding to fully incoherent (red) and fully coherent (blue) illumination. A gray rectangle marks the area where the centered cosine similarity is below 0.9. The dashed line indicates where $d \approx d_i$. The \textit{x}-axis is at log-scale. The results are averaged over 5 different digits, where for each digit we used 5 different random phase masks. (c) The same as in (b) only for an imaging system, \textit{i.e.} where the diffractive element is a lens with quadratic phase that preforms unit imaging up to rotation. 
   (d) Several scenarios where $d \approx d_i$ and $a \approx D$, in which the system inherently does not operate in the incoherent regime.}
   \label{fig:limits}
\end{figure}

Consider an optical system, as in Figure~\ref{fig:limits}a (but with circular rather than square pupils), comprising a planar object, a thin diffractive element of diameter $D$ placed a distance $d_i$ from the object, and an output plane. 
It is known that such a system can be considered as operating under spatially incoherent illumination if the spatial coherence length at the object plane satisfies \cite{goodman2015statistical} (see \ref{sm:limits})
\begin{equation}\label{eq:limits}
    l_c \le \frac{1}{\sqrt{\pi}} \cdot \frac{1}{f^{\text{sys}}_{\max} + f^{\text{obj}}_{\max}},
\end{equation}
where $f^{\text{sys}}_{\max} =\frac{D}{2\lambda d_i}$ is the maximal spatial frequency preserved by the system and $f^{\text{obj}}_{\max}$ is the maximal spatial frequency of the object's transmittance function (\textit{i.e.} assuming the Fourier transform of the object's transmittance is supported on a disk of radius $f^{\text{obj}}_{\max}$). If the condition in Eq.~\eqref{eq:limits} is not met, then the system does not behave as if the object emits spatially incoherent light. In such a case, the partial coherence of the illumination impinging on the object may need to be taken into consideration. In \ref{sm:limits} we show that this condition characterizes not only imaging systems, but also any system that is approximately shift-invariant.
 
Equation~\eqref{eq:limits} implies that a system \emph{cannot} be considered as operating under completely incoherent illumination when the coherence length $l_c$ at the object plane is larger then the minimal feature size preserved by the system, $S_F=(f^{\text{sys}}_{\max})^{-1}=\frac{2\lambda d_i}{D}$. Interestingly, this is often the case. Indeed, let us assume that the object is illuminated by a planar incoherent source of diameter $a$, placed a distance $d$ from the object. In this case, the coherence length on the object is $l_c = \frac{\lambda d}{\sqrt{\pi}a/2}$ (see Section~\ref{sec:framework}) and is therefore larger than $S_F$ when
\begin{equation}\label{eq:condition}
\frac{a}{d}<\frac{D}{d_i}.
\end{equation}
Namely, when the numerical aperture (NA) of the illumination is smaller than the collection NA of the system, the commonly used assumption that the object emits spatially incoherent light, is inaccurate. 
This occurs in numerous applications, particularly in reflection settings involving active illumination by incoherent sources (\textit{e.g.} LED). Indeed, when the source is located near the entrance pupil of the optical system, we have that $d \approx d_i$, and if the source dimensions are similar to those of the entrance pupil, then $a \approx D$. This setting is inherent to \textit{e.g.} autonomous vehicles, indoor drones, smartphones, and active virtual reality headsets (see Figure~\ref{fig:limits}d). It is also inherent to reflected light microscopy \citep{edwards2014effects}, as in most commercial microscopes that use epi-illumination, the light illuminating the sample passes through the same NA as that collected from the sample. Thus, when the condenser's NA is open to the maximal extent, the spatial coherence length on the imaged sample equals the minimal feature size $S_F$.

It is important to note that the discussion above is relevant to imaging and non-imaging systems alike. Namely, numerous real-world imaging systems do not operate in the incoherent regime. However, for imaging systems, violating the incoherence assumption typically does not lead to a markedly different intensity pattern at the output, while for non-imaging systems, the intensity pattern at the output may strongly depend on the coherence level. This is numerically illustrated in Figure~\ref{fig:limits}. Here, we simulate the system shown in Figure~\ref{fig:limits}a, which comprises an incoherent square source of side length $a$, located a distance $d$ from the object plane. The object is located a distance $d_i$ from a square diffractive element of side length $D$. 
The spatial coherence degree at the object plane is controlled by changing the distance $d$ between the incoherent light source and the object plane (see Section~\ref{sec:framework} and \ref{sm:partially}). The object is a handwritten digit of dimensions 0.84 $\times$ 0.84 mm. The source and the diffractive element dimensions are 0.84 mm $\times$ 0.84 mm. Therefore $a=D$, which following Eq.~\eqref{eq:condition}, results in $l_c=S_F$ when $d=d_i$.

We simulated both an imaging system and a diffractive network. For the latter, we took the diffractive element to be a random phase mask. Its pixels, each of dimensions 10 $\mu$m $\times$ 10 $\mu$m, were independently drawn at random between $0$ and $2\pi$. For the imaging system, we simulated a lens with quadratic phase that performs unit imaging up to rotation. We calculated the intensity patterns at the output plane of this optical system for 35 different $d$ values, that translate into coherence lengths $l_c$, ranging from 800 $\mu$m to 2 $\mu$m. All the numerical experiments described above were simulated using monochromatic illumination of $\lambda=550$ nm. Several example output intensity patterns corresponding to different spatial coherence conditions are displayed in Figure~\ref{fig:limits}b,c, for random and quadratic phases, respectively. 

In order to quantify the difference between the intensity patterns resulting from different spatial coherence conditions we focused on the two extreme cases, where the illumination is (i) coherent and (ii) incoherent. We calculated the centered cosine similarity between (i) and all intensity patterns and between (ii) and all intensity patterns, for both the random and the quadratic phases. As can be seen on the right of Figure~\ref{fig:limits}b,c the centered cosine similarity for the imaging system is relatively high for all coherence conditions, as expected. A black dashed line indicates where $d=d_i$. However, for the diffractive network, the centered cosine similarity decreases as the coherence conditions draw apart from the ones used for (i) or (ii). Thus, for diffractive networks there is a substantial regime, marked in gray, where the output intensity of the network is markedly different from both the coherent and the incoherent cases. It can also be seen from the centered cosine similarity score in this area, which is below 0.9 for both the coherent and incoherent limits. See \ref{sm:figure-details} for more details about the optical system and the mentioned calculations. 

Figure~\ref{fig:limits}d shows examples for applications that involve reflectance. In all these cases the illumination source is located near the camera's entrance pupil and is of similar dimensions. Therefore, in order for a diffractive network to be deployed in any of these, or similar, applications, the partial coherence of the light has to be taken into consideration.
    
\subsection{Coherence aware framework for diffractive neural networks}
\label{sec:framework}

To account for partially incoherent illumination, we propose a coherence-aware framework for training diffractive neural networks. Our approach relies on the van Cittert–Zernike theorem \citep{van1934wahrscheinliche, zernike1938concept}, which describes the relation between the intensity pattern of a planar incoherent source and the coherence function at some observation plane.

\begin{figure}[t]
  \centering
   \includegraphics[width=\linewidth]{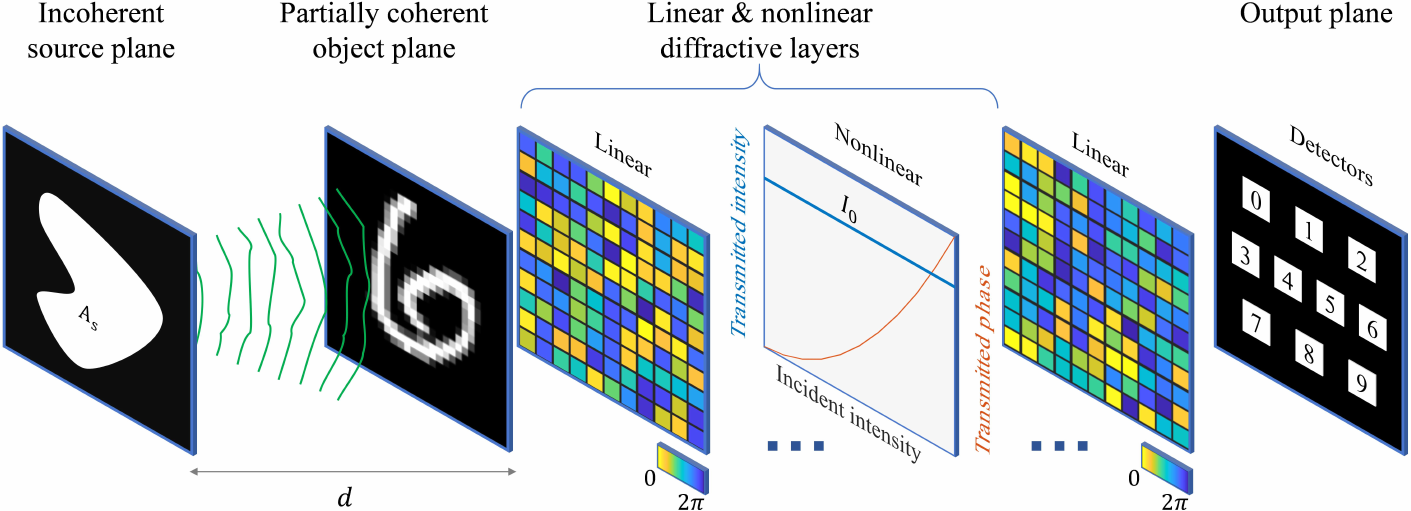}
   \caption{\textbf{Optical configuration for coherence-aware networks}. A planar, incoherent, uniformly bright source of area $A_S$ is placed a distance $d$ from the object plane. The parameters $A_S$ and $d$ control the degree of spatial coherence at the object plane. As opposed to networks for fully coherent and fully incoherent light, here computing the intensity at the output plane necessitates propagating fields from the source plane rather than starting at the object plane. The figure presents an example of a classification network comprising power-preserving linear and nonlinear diffractive layers. The output plane contains an array of detectors, one for each object class. In this setting, the network may be optimized to concentrate light on the detector corresponding to the predicted class.}
   \label{fig:source}
\end{figure}

We consider the optical setup depicted in Figure~\ref{fig:source}, where a planar uniformly bright source of area $A_s$ is located a distance $d$  from the object plane. This setup allows controlling the degree of spatial coherence on the object by tuning $A_s$ and $d$. Specifically, according to the van Cittert–Zernike theorem, for a monochromatic source of wavelength $\lambda$, the coherence area $A_c$ on the object plane is given by \citep{goodman2015statistical}
\begin{equation} \label{eq:coherence_area}
    A_{c}=\frac{(\lambda d)^2}{A_{s}}.
\end{equation}

In~\ref{sm:partially} we discuss how the coherence area can be computed for broadband sources. Thus, for a fixed source area $A_s$, as the distance to the object increases, the illumination on the object plane becomes more coherent (\textit{i.e.}~has a larger coherence area $A_c$). It is convenient to use the notion of coherence length, defined as $l_c=\sqrt{A_c}$. Here, we employ a square light source with side length $a$, so that $A_{s}=a^2$ and thus the coherence length is $l_c=\frac{\lambda d}{a}$. 
As opposed to the fully coherent and fully incoherent settings, in which the optical system is commonly regarded as starting at the object plane, in our setting it is important to simulate the system from the source plane to the output plane where the detector is placed. Consider for simplicity a diffractive network comprising only linear elements (we later extend the discussion to nonlinear settings). Let $H(x,y,x',y',\lambda)$ denote the field impulse response from the source to the detector, where $(x,y)$ and $(x',y')$ are the coordinates in the source and output planes, respectively, and $\lambda$ is the wavelength. $H(x,y,x',y',\lambda)$ has dimensions of inverse length squared. Since the source is spatially incoherent, it can be regarded as an extended collection of independent radiators. For each wavelength, the field from each individual radiator coherently propagates through the optical system, and the resulting intensity at the output plane is the integral over the intensities corresponding to the different radiators. Namely, we have that  
\begin{equation}\label{eq:H}
    I_{\text{out}}(x',y',\lambda) = \kappa \iint |H(x,y,x',y',\lambda)|^{2}\, I_{\text{source}}(x,y,\lambda)\,dx dy,
\end{equation}
where $I_{\text{source}}(x,y,\lambda)$ and $I_{\text{out}}(x',y',\lambda)$ are the optical intensities per wavelength at the source and output planes, respectively. The constant $\kappa$ has dimensions of length squared. The intensity over the entire spectrum is obtained by integrating over the wavelength, 
\textit{i.e.}, $I_{\text{out}}(x',y') =  \int I_{\text{out}}(x',y',\lambda) \,d\lambda$.
\begin{figure}[t]
  \centering
   \includegraphics[width=\linewidth]{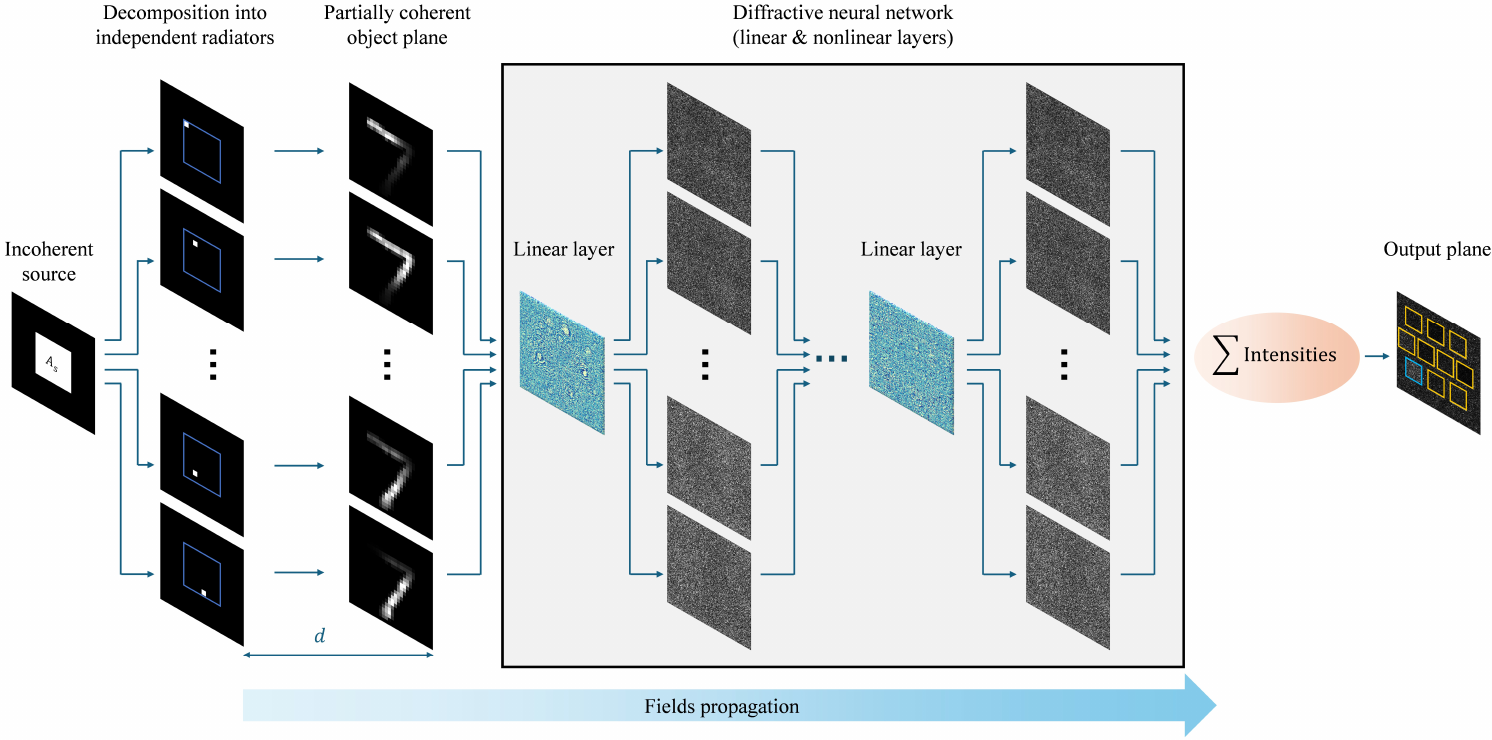}
   \caption{\textbf{The proposed method for simulating a network with partially spatial coherent light.} A spatial incoherent square light source is located at a distance $d$ in front of an object plane. Each pixel (radiator) in the light source is separately propagated and multiplied by the object's amplitude. Each multiplication result is separately propagated through the diffractive network. The intensity is calculated in the output plane for each of the light source's pixels and summed for all of them. The resulting intensity in the output plane is summed in pre-determined areas, marked by yellow squares. The area with the highest intensity is chosen as the correct classification, marked with a cyan square.} 
   \label{fig:method}
\end{figure}
As depicted in Figure~\ref{fig:method}, to computationally implement Eq.~\eqref{eq:H}, we coherently propagate the field emitted from each pixel of our square light source a distance $d$, multiply it by the object's amplitude, and then continue to propagate it through the rest of the diffractive network, up to the output plane. The overall intensity at the output plane is obtained by summing up the intensity images contributed by all the pixels in the source.
Our framework facilitates the incorporation of optical nonlinear layers, which are crucial for unlocking the full potential of neural networks \citep{wetzstein2020inference, mengu2022intersection, mengu2019analysis}. These layers, which serve the roles of the nonlinear activation functions used in digital neural networks, do not introduce any learnable parameters into the optimization process. To accommodate such layers, we compute the intensity patterns of the incoming light, consider transmission through the layer based on its nonlinear function, and then proceed to propagate the fields originating from individual points on the source through the subsequent linear layers up to the next nonlinear layer.

It is instructive to note that a diffractive network operating under partially incoherent light is conceptually different from standard digital feed-forward networks, as well as from diffractive networks operating under coherent light. Indeed, in digital feed-forward networks, each layer operates on the output of the preceding layer. This is also the case for diffractive networks operating under coherent light, where the field at the output of each layer depends only on the field at its input. In contrast, when operating under partially coherent illumination, it is necessary to account not only for the field at the input of each layer but rather also for the coherence between every two points in its plane. Thus, as opposed to the coherent setting, training a network for partially coherent light necessitates accounting for the geometry of the light source and its distance from the object. 

\begin{figure}[t]
  \centering
   \includegraphics[width=\linewidth]{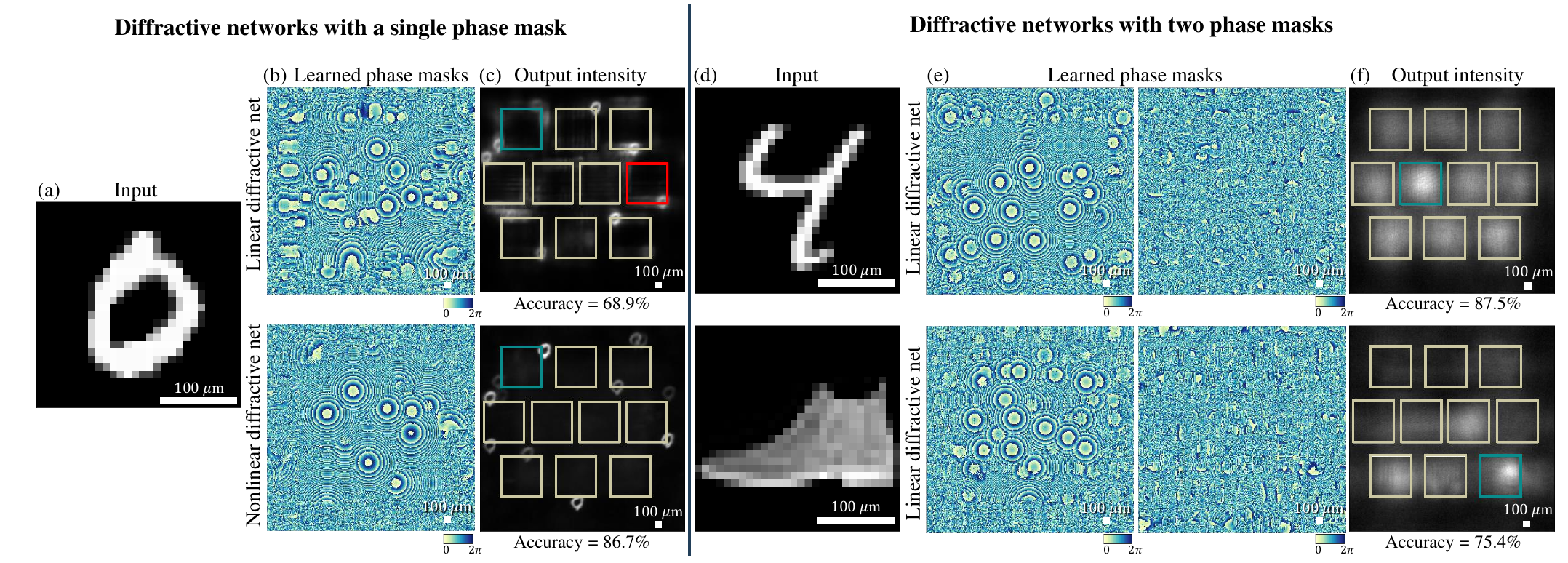}
   \caption{\textbf{Versatility of the proposed framework.} The left-hand side of the figure depicts diffractive networks with a single phase mask, both without (top) and with (bottom) a nonlinear element placed right after it. The right-hand side of the figure depicts diffractive networks with two linear phase masks. The networks are trained for classification of handwritten digits and fashion items. (a),(d) Examples of items for classification in the object plane. (b),(e) The learned phase masks of the linear and nonlinear diffractive networks. (c),(f) Illustrations of the output intensity patterns indicating the network's classification predictions. The cyan square corresponds to the class of the ground-truth digit. In all cases except for the top pane in (c), the networks' predictions are correct. In the top pane of (c), we show an example of an incorrect prediction of a linear network, indicated by a red square.}
   \label{fig:incoherent_classification2}
\end{figure}

In Figure~\ref{fig:incoherent_classification2} we demonstrate the versatility of our approach, showcasing its applicability for training linear and nonlinear diffractive networks with one and two phase masks. We focus on classification of handwritten digits from the MNIST dataset \citep{lecun1998mnist} and fashion items from the Fashion-MNIST dataset \citep{xiao2017fashion}. The networks were trained for classifying objects of size $280$ \mbox{$\mu$m $\times$ $280$ $\mu$m} with diffractive layers of size $3$ mm $\times$ $3$ mm, each containing a grid of $300\times 300$ pixels. The output plane dimensions were also \mbox{$3$ mm $\times$ $3$ mm}. The object plane, the diffractive layers and the output plane were equally spaced with a $5$ cm spacing. In these numerical experiments we used a light source with flat spectrum in the range \mbox{$400$ nm - $700$ nm} and set $d=0$ to obtain spatially incoherent illumination. 
The networks were trained to focus light on designated \mbox{$600$ $\mu$m $\times$ $600$ $\mu$m} regions at the output plane, each corresponding to a different object class (Figure~\ref{fig:incoherent_classification2}c,f). Example input images are shown in Figure~\ref{fig:incoherent_classification2}a,d and the learned phase masks are presented in Figure~\ref{fig:incoherent_classification2}b,e. The masks are characterized by circular features scattered across the plane, serving as lenslets to focus light to the next layer. The reported network output corresponds to the region with the highest intensity (integrated over the entire region). The reported accuracies were measured on the entire test sets. The images in Figure~\ref{fig:incoherent_classification2}c are examples of incorrect (top) and correct (bottom) classification results.
For the nonlinear network, we used a single nonlinear layer in the form of a thin film of photorefractive crystal (SBN:60) \citep{waller2012phase, christodoulides1997theory}, following previous works \citep{yan2019fourier, dou2020residual}. Digital implementation and training details can be found in Section \ref{methods}. Further discussion and analysis of diffractive networks for incoherent illumination can be found in~\ref{sm:incoherent light}.

\subsection{Image classification using partially coherent illumination}
\label{res_partially}
We next investigate the effect of spatial and temporal coherence on the performance of diffractive networks. We focus on two-layer linear diffractive networks for image classification with the same physical dimensions as mentioned above. Here we report results for the MNIST dataset. Results for the Fashion-MNIST dataset can be found in \ref{sm:partially}. We start by numerically examining the nonblind case in which the coherence parameters are precisely known at train time. We then proceed to study coherence-blind networks, which are oblivious to the precise levels of coherence. Our coherence-blind training is similar to works that trained diffractive networks for varying levels of layer-to-layer misalignment~\cite{mengu2020misalignment} and to changes in the size, orientation and distance of the object from the network~\cite{mengu2020scale}.

\begin{figure}
  \centering
   \includegraphics[width=\linewidth]{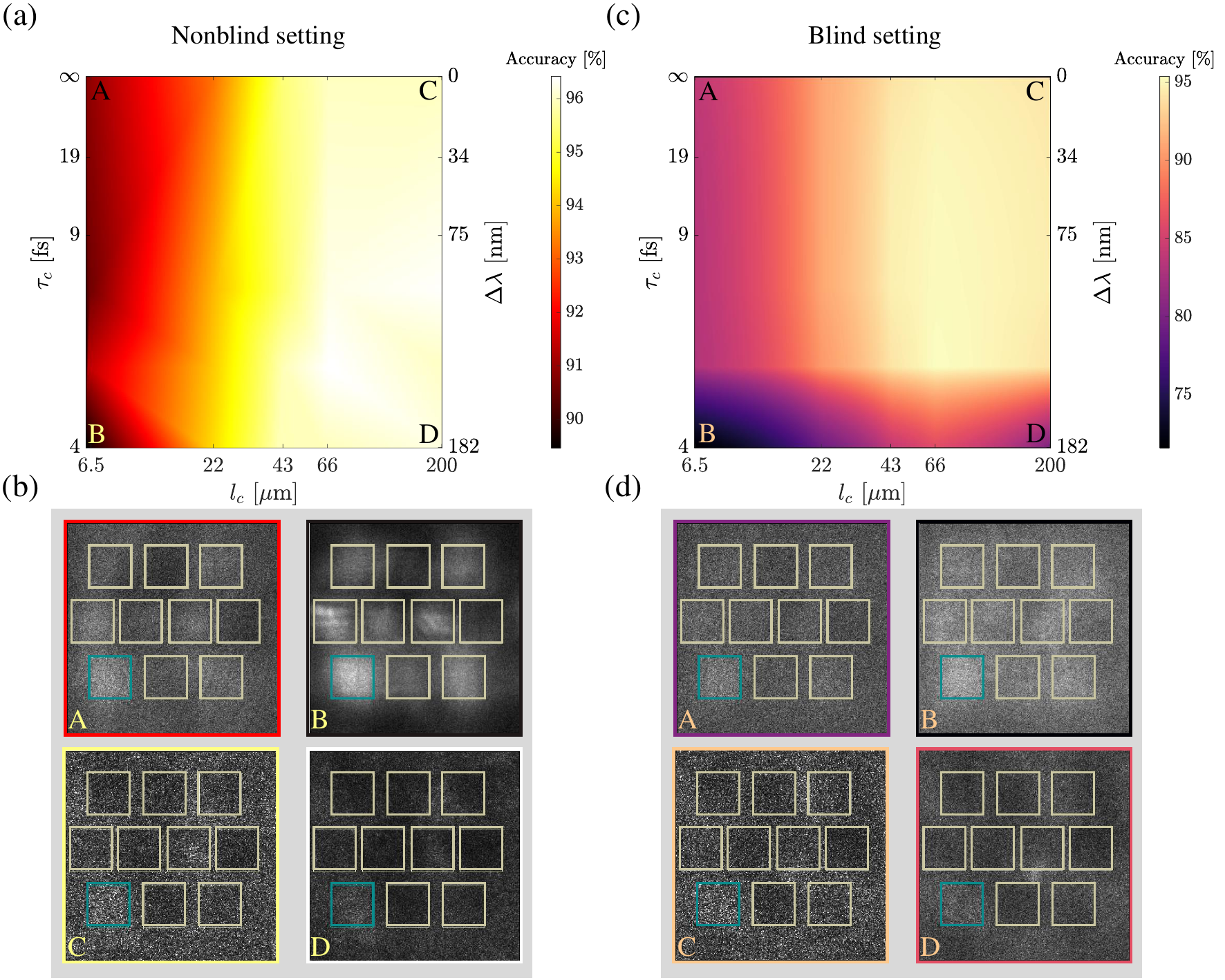}
   \caption{\textbf{Effect of spatial and temporal coherence on the performance of diffractive networks.} The top part of the figure shows the accuracy rates for different levels of spatial and temporal coherence, for nonblind (a) and blind (c) settings. The horizontal axis measures spatial coherence via coherence length. The vertical axis measures temporal coherence and has two scales, one for coherence time (left) and another for the associated wavelength bandwidth around the central wavelength of 550 nm (right). Both axes are on a logarithmic scale. The different colors on the graph indicate the resulting accuracy. The edges of each graph indicate different spatially and temporally coherent states, from the temporally and spatially incoherent B to the temporally and spatially coherent C. The bottom part of the figure displays the intensity patterns at the output plane for the nonblind (b) and blind (d) settings. The images are shown for the same input image from the test set (a digit 7), for the edge cases marked by A, B, C, and D in (a) and (b).}
   \label{fig:partial}
\end{figure}

Figure~\ref{fig:partial} presents the results for both settings. In these numerical experiments, we set the source to be a square of side length $120$ $\mu$m, and controlled the spatial coherence by choosing~$d$ from among five values corresponding to coherence lengths of approximately $l_c = \{6.5, 22, 43, 66, 200\}$~$\mu$m. We used a source with a Gaussian spectrum centered at a wavelength of $550$ nm and controlled the temporal coherence by tuning the width of the Gaussian (see \ref{sm:incoherent light} for experiments with a flat spectrum). Specifically, we used wavelength bandwidths of $\Delta\lambda=\{0, 34, 75, 182\}$ nm corresponding to coherence times of $\tau_c = \{\infty, 19, 9, 4\}$ fs. 
More information about the coherence properties and the sampling of the wavelengths is provided in~\ref{sm:partially} and \ref{sm:incoherent light}, and a discussion about the effect of phase quantization, which may be introduced in fabrication, is provided in~\ref{sm:quant}. 
Figure~\ref{fig:partial}a shows the classification accuracy rates for the nonblind case. Here, we trained~20 different networks, one for each combination of spatial coherence~$l_c$ and temporal coherence~$\tau_c$. We evaluated each of them on the entire test set, using the same coherence parameters on which they were trained. These results serve as an upper bound on the performance expected in practical settings, where the coherence properties at test time may not precisely match those used during training. In this case the accuracy rates range from 89.3\% to 96.4\%, and depend more strongly on the degree of spatial coherence than on the degree of temporal coherence. 
Specifically, for almost any level of temporal coherence, the results improve as the spatial coherence grows larger, where the biggest variations occur for temporally incoherent light.
Figure~\ref{fig:partial}b shows the intensities at the output plane for the four coherence settings marked A,B,C,D in Figure~\ref{fig:partial}a, all for the same input image, a digit 7 from the test set. For incoherent light (setting B), which corresponds to the lowest classification accuracy, the intensity pattern is smooth. As the spatial and/or temporal coherence levels increase (settings A,C,D), the intensity images become dominated by speckle patterns, which the networks exploit for achieving better classification accuracy.

Figures~\ref{fig:partial}c,d show the results achieved by a single coherence-blind network, which was trained on all coherence settings simultaneously. Specifically, we trained a network by randomly choosing in each training batch one combination of $l_c$ and $\tau_c$, and updating the phase masks only with respect to those settings. It should be noted that the extreme range of coherence levels we consider, encompasses sources from laser to sunlight and LEDs, and beyond. Such uncertainty in the coherence conditions is expected to exceed that encountered in any realistic application. Therefore, these results can be considered as a lower bound for the accuracy that can be expected in practical settings. 
The accuracy rates range from 71.5\% to 95.4\%, and as in the nonblind setting, they depend more strongly on the degree of spatial coherence. However, as opposed to the nonblind setting, where the network's accuracy depends mostly on the spatial coherence, in the blind case the network's performance depends also on the temporal coherence, where the results generally improve as the light's bandwidth decreases.
\begin{figure}
  \centering
   \includegraphics[width=\linewidth]{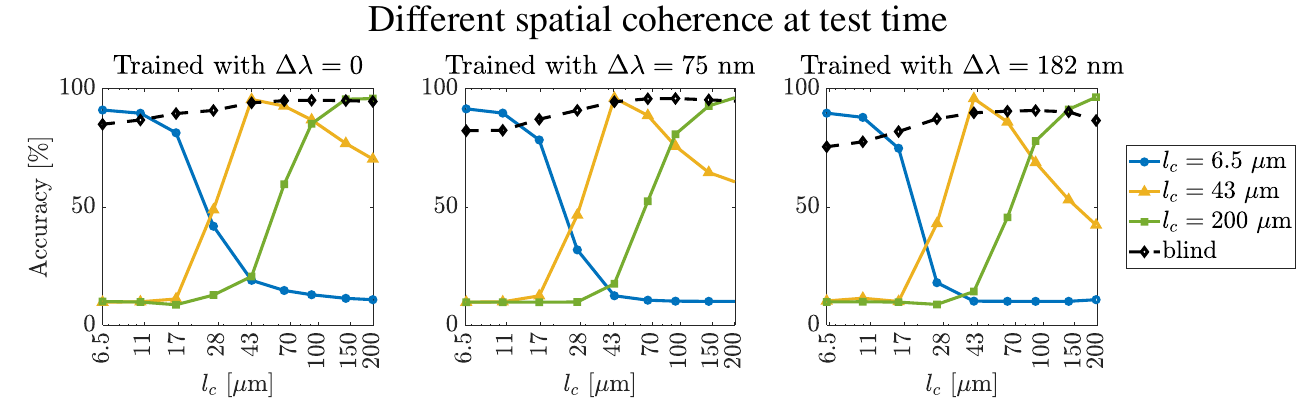}
   \caption{\textbf{The effect of mismatch in spatial coherence between training time and test time.} Each plot displays the results of three different diffractive networks, all trained with the same level of temporal coherence (specified in the title) and with different levels of spatial coherence (specified in the legend). For reference, each plot also depicts a spatial coherence blind network as a black dashed line. This spatial coherence blind network was trained with the same level of temporal coherence as the nonblind networks.}
   \label{fig:s_test}
\end{figure} 

Since training on a large range of coherence levels results in a notable drop in accuracy, it may be desirable to train only on a narrow range around the assumed coherence. For example, one may want a diffractive network that is blind only to the spatial coherence level or only to the temporal coherence level (corresponding to training over only a horizontal or only a vertical cross section of the coherence settings in Fig.~\ref{fig:partial}c). We exemplify blindness to only spatial coherence in Fig.~\ref{fig:s_test} and to only temporal coherence in Fig.~\ref{fig:t_test} in \ref{sm:partially}. Specifically, in Figure~\ref{fig:s_test} the black dashed curves show the accuracy of a coherence-blind network that was \emph{trained for a range of spatial coherence lengths and a single temporal coherence length}. As opposed to Figure~\ref{fig:partial}c, here there is a negligible drop in accuracy with respect to the nonblind networks of Figure~\ref{fig:partial}a.

In practical use cases, the coherence properties at deployment time may deviate from those used at training. For example, the spatial coherence of sunlight can vary with weather conditions \citep{divitt2015spatial}. It is therefore instructive to assess the effect of mismatch between the coherence levels used in training and those used at test time. Figure~\ref{fig:s_test} further shows the performance of diffractive networks from the simulation of Figure~\ref{fig:partial}a, which had been trained for specific spatial and temporal coherence levels, when tested on different spatial coherence conditions. 
In each plot in Figure~\ref{fig:s_test}, three networks (colored curves) are tested with a different fixed temporal coherence and a varying spatial coherence. Each network had been trained with the correct temporal coherence (specified in the title) and a different level of spatial coherence (specified in the legend). The performance peaks when the spatial coherence used at test time matches the spatial coherence used at train time, and decreases gradually as the coherence mismatch grows larger. For large mismatches, the classification accuracy drops to 10\%, which corresponds to a random guess. 

We also investigated the effect of mismatch in temporal coherence between training time and test time. We observed that diffractive networks tend to exhibit greater tolerance to mismatches in the level of temporal coherence than to the level of spatial coherence. For further discussion see \ref{sm:partially}.

\subsection{Application to classification of phase objects}
\label{sec:blood}

Biological cells and tissues are known as phase objects because they primarily affect the phase of incident light, rather than its amplitude. This makes them largely transparent to conventional optical systems. To image them, it is common to convert them into amplitude objects using staining or fluorescent tags. Alternatively, label-free quantitative phase imaging methods can directly sense these phase objects by recording interferometric measurements, which are then computationally processed to reconstruct the phase information.~\cite{park2018quantitative}.
Diffractive networks were recently proposed as an alternative all-optical method for quantitative phase imaging~\cite{mengu2022all} and for other different tasks that involve phase objects, such as classification of overlapping phase objects~\cite{mengu2022classification}.  

Here, we numerically demonstrate our coherence aware framework's application to classification of phase objects. 
As mentioned in Section~\ref{sec:effect_on_diff}, in reflected light microscopy with equal illumination and collection NAs, the minimal feature preserved by the system, $S_F$, is the same as the spatial coherence length on the object, $l_c$. 
Inspired by this setting, where our coherence-aware diffractive network is of crucial importance, we simulated an optical system with equal illumination and collection NAs ($0.09$) resulting in $S_F=l_c=5.8$ $\mu$m when using a spacing of $d_1=1.6$ cm and phase masks of $3$ mm side length. 

\begin{figure}
  \centering
   \includegraphics[width=\linewidth]{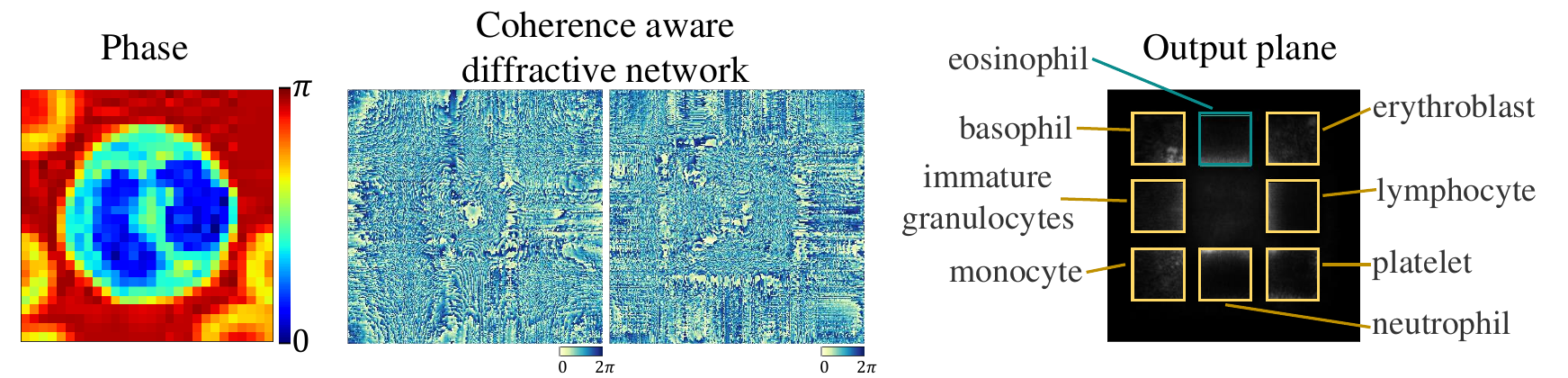}
   \caption{\textbf{Coherence-aware network for phase objects classification.}
    A coherence-aware diffractive network is used to optically process phase objects. Before network processing, the input images exhibit constant amplitude and varying phase information, which is invisible to standard cameras. The network is designed to classify eight distinct types of transparent red blood cells based on their phase profiles. The network's output plane is divided into pre-determined regions, highlighted in yellow, where the region with the highest intensity indicates the network's classification (marked with a cyan square).}
   \label{fig:reflected}
\end{figure}

We used the BloodMNIST datasets from the MedMNIST collection~\cite{medmnistv2, acevedo2020dataset}, which contains images of blood cells, organized into 8 classes, by different types of blood cells.  
Each image was converted into a phase image by converting the image values to the range $[0,\pi]$ and using these values as the phase of an object with a constant amplitude of~1. 
The proposed network processed these phase elements using its diffractive elements. The network's output is an intensity image, divided into pre-determined regions, where the region with the highest intensity is declared as the predicted class.
We trained a coherence aware network using this dataset, with the spatial coherence condition described above ($l_c=5.8$ $\mu$m) and temporally coherent illumination (a single wavelength). The classification accuracy of this network is 88.2\%.  
Figure \ref{fig:reflected} illustrates an example of the phase objects we aim to classify (left), the resulting phase masks from the network training process (middle) and the associated output plane (right). 

To illustrate the importance of coherence awareness, we further trained a diffractive network under the incorrect assumption of spatially (and temporally) coherent illumination. When this network is evaluated with the actual spatial coherence conditions, as stated above, it achieves a classification accuracy of only 25.8\%. This is despite the fact that it achieves 94.3\% classification accuracy when tested with its assumed (incorrect) coherence level. 
Similarly, we trained a diffractive network under the incorrect assumption of spatially incoherent (and temporally coherent) illumination. When evaluating this network with the correct coherence conditions, its classification accuracy is only 31.4\%, despite achieving an accuracy of 80.7\% under its assumed (incorrect) setting. 
This deterioration due to the mismatch in the spatial coherence conditions between training and testing, highlights the importance of coherence awareness in practical settings.

\section{Discussion}
\label{discussion}
We demonstrated the importance of coherence awareness in diffractive neural networks and proposed a general framework for optimizing a diffractive network with any prescribed degree of spatial and temporal coherence illumination. We additionally proposed a method for training coherence-blind networks and numerically demonstrated their superiority over nonblind networks when the coherence conditions at deployment deviate from those assumed during training. Our framework provides means for adapting diffractive networks to the environmental illumination rather than restricting their use to particular specialized settings, and constitutes a first step towards understanding the potential and limitations of using optical computing with uncontrolled environmental lighting.

Our framework can use any technique for simulating the propagation of the incoherent light emanating from the source. In our simulations, we coherently propagated each point on the source through the entire network, and summed the resulting intensities at the output plane. An alternative method, known as modal expansion \citep{wolf1982new}, relies on coherently propagating many random-phase patterns on the source plane and averaging the resulting intensities at the output plane. This approach has been recently used for training diffractive networks for spatially incoherent illumination \citep{rahman2023universal, yang2024complex}. In \ref{sm:modal expansion} we compare between the approaches for incoherent illumination, and find that our technique is often advantageous for diffractive networks with more than two layers.

It is insightful to note that in linear networks, the relation between the transmittance function of the object and the output intensity is always quadratic (see Eq.~\eqref{eq:space-inv-intensity} in \ref{sm:limits}), regardless of the level of spatial or temporal coherence. However, in the extreme case where the light is completely spatially incoherent, the network's output intensity also becomes linear in the input intensity. In our setting, the input intensity is roughly equal to the object's transmittance function, because our objects are nearly binary. This implies that in our setting, the network becomes linear under spatially incoherent light. This may explain the behavior seen in Fig.~\ref{fig:partial}, where the performance degrades as the light becomes more spatially incoherent. Please see further discussion about this point in \ref{sm:nonlinearity_partial}.

Recent work has explored the use of optoelectronic nonlinear activation functions within optical neural networks, focusing on broadband spatially incoherent illumination \citep{wang2023image}. An important future research direction is to study the effect of the degree of coherence on such systems. This would allow for characterizing and optimizing their performance under changing environmental illumination conditions.

More broadly, while the effects of light coherence have been extensively studied in traditional scenarios such as imaging or other optical configurations with gradual changes of phase accumulation through its elements, there remains significant unexplored territory in modern optical systems. These systems, which typically leverage advanced nanofabrication and material synthesizing techniques, introduce abrupt changes in the optical response of their constituent elements. Specific examples include optical metasurfaces for imaging and active wavefront control, as well as luminescent solar concentrators for enhancing solar energy harvesting by capturing radiation over large areas. In the case of optical metasurfaces, research has mainly focused on manipulating coherent laser radiation. Conversely, solar concentrators unavoidably operate with broadband incoherent sunlight radiation. Both examples could benefit from more exploration of light manipulation in the intermediate regime of partially coherent light. We believe that our findings, in addition to paving the way for the adoption of all-optical neural networks in real-world applications using natural light exclusively, will also promote further research in these areas and shed light on the crucial role of coherence in contemporary optical systems and their practical implementations.

\section{Materials and methods}
\label{methods}

\subsection{Nonlinear layer}
For the nonlinear networks, we modeled a thin film of photorefractive crystal (SBN:60) \citep{waller2012phase, christodoulides1997theory} as the all-optical nonlinear layer, according to previous works \citep{yan2019fourier, dou2020residual}. The nonlinearity is manifested by a change in the refractive index as a function of the light intensity, thus affecting solely the phase of the light and preserving its power. This layer serves the role of a nonlinear activation function in the diffractive network. In our simulations, we positioned these layers immediately after a diffractive layer, without free space propagation in between. The physical parameters of the photorefractive crystal were taken from the work of Yan et al.~\citep{yan2019fourier} Details are available in the supplemental material of their work.

\subsection{Digital implementation and training details}
For training our diffractive networks, we used a dataset $\mathcal{D}$ of images, $x_i$ and their corresponding labels $y_i$, $\mathcal{D}=\{(x_i,y_i)\}_{i=1}^N$. Each image $x_i$ was zero-padded to the size of the diffractive layer, which is $300\times 300$ pixels. The output plane dimensions were also $300\times 300$ pixels. We trained our diffractive networks for multi-class image classification via backpropagation \citep{rumelhart1986learning}, using the cross entropy loss between the network's predictions and the ground truth labels.

Free space propagation between the layers was simulated using the angular spectrum method~\citep{goodman2005introduction} and the Rayleigh-Sommerfeld diffraction formulation. All fields were adequately sampled and zero-padded to prevent aliasing. See~\ref{sm:forward_prop} and \ref{sm:digital} for more details. 

We used the MNIST \citep{lecun1998mnist} and Fashion-MNIST \citep{xiao2017fashion} datasets, which contain $28\times 28$ grayscale images of handwritten digits and fashion items, respectively, each belonging to one of 10 classes. As training is time consuming, we used only a representative subset of 5000 images from the datasets for training ($\sim500$ from each class). We evaluated our trained diffractive networks on the entire test set, which consist of 10000 images ($\sim1000$ per class) that are not a part of the training set. We trained our diffractive networks with the Adam optimizer \citep{kingma2015adam} with a learning rate of $\eta=1\times10^{-2}$ for 50 epochs and with an effective batch size of 32. We used a training scheduler which reduces the learning rate by a factor of 10 when the training loss stops improving for 5 consecutive epochs. All the numerical experiments described in this paper were executed with those hyperparameters, unless explicitly noted otherwise. 
The simulations were implemented using the Pytorch deep learning framework \citep{paszke2019pytorch} and were executed on a Linux machine with NVIDIA GeForce GTX 1080 Ti GPU. 
Training of a diffractive network with two layers and the hyperparameters mentioned above requires approximately 4 GB of memory and takes approximately 25 hours to complete on that machine.

For classification of phase objects (Section~\ref{sec:blood}) we used the BloodMNIST datasets from the MedMNIST collection~\cite{medmnistv2, acevedo2020dataset}.
This dataset contains $28\times 28$ color images of red blood cells, organized into 8 classes. The images were first converted to graysacle and then to phase objects. 
This dataset contains $\sim 12000$ training images and $\sim 3500$ test images.
We employed the same training procedure mentioned above and the entire training set to train networks in this setting. The networks were evaluated on the entire test set.

\subsection{Efficient training of linear networks for spatially incoherent light}\label{sec:efficient_incoherent}
In the specific case of spatially incoherent illumination, where $d=0$, the source plane and the object plane coalesce. In this situation, the source's dimensions should match the object’s dimensions to ensure the entire object is properly illuminated. Therefore, for spatially incoherent illumination, the straight-forward implementation of our approach requires propagating through the network each pixel of each image within the batch. For a 32-image batch of the MNIST dataset, this boils down to propagating 25088 pixels through the network in each batch. However, this naive strategy can be significantly improved upon.

First, instead of separately propagating each pixel of the source, it is only required to propagate only the nonzero pixels. The number of nonzero pixels in an average 32-image batch from the MNIST dataset is $\sim$5700, thus requiring only $\sim$23\% of the computations of the naive approach. Second, for linear diffractive networks, it is possible to reduce the computational load even further by exploiting the fact that the intensity pattern at the output plane is a linear combination of the intensity patterns resulting from each of the nonzero pixels in the object plane. Thus, by calculating only once the output intensity patterns of all the possible nonzero pixels in a given batch, we can calculate the intensity pattern of any object in this batch. In an average 32-image batch from the MNIST dataset the nonzero pixels are located in $\sim$700 different locations. Thus, by calculating only the intensity pattern for each pixel in each of these $\sim$700 locations, we can calculate the intensity patterns for the entire batch, by using only 3\% of the forward passes of the naive approach.  

\clearpage

\bibliography{references}

\section*{Acknowledgements}
The research of T.M.~was partially supported by the Israel Science Foundation (grant no.~2318/22).

\clearpage

\doublespacing

\renewcommand{\thesection}{Supplementary Note \arabic{section}}\setcounter{section}{0}  
\renewcommand{\thesubsection}{\arabic{section}.\arabic{subsection}}

\renewcommand\thefigure{S\arabic{figure}}  
\setcounter{figure}{0}  
\renewcommand\theequation{S\arabic{equation}}  
\setcounter{equation}{0}  

\clearpage
\section{Further details about Figure~\ref{fig:limits}}
\label{sm:figure-details}
We simulated both an imaging system and a diffractive network to achieve the results illustrated in Fig.~\ref{fig:limits}. For the imaging system we simulated a lens with quadratic phase. The lens transfer function is
\begin{equation}
    t_{\text{lens}}(x,y)=\exp\left\{-j\frac{\pi}{\lambda f}(x^2+y^2)\right\},
\end{equation}
where $(x,y)$ are the lateral coordinates in the lens' plane. 

For the diffractive network we took the diffractive element to be a random phase mask, that is similar in nature to the learned phase masks of diffractive networks \cite{Lin2018, mengu2019analysis}. The transfer function of this diffractive layer is
\begin{equation}    
     t_{\text{layer}}(x,y)=\exp\{-j\phi(x,y)\},
\end{equation}
where $\phi(x,y)$ is the random phase and $(x,y)$ are again the lateral coordinates. 
The free space propagation between all optical elements was performed using the Rayleigh-Sommerfeld diffraction formulation and the angular spectrum method~\citep{goodman2005introduction} (see \ref{sm:forward_prop}).  

We used the centered cosine similarity to calculate the similarity between the output intensity patterns. For any pair of intensity images, flattened into column vectors $I_1$ and $I_2$, the centered cosine similarity is defined as
\begin{equation}
    \frac{(I_1-\bar{I}_1)^{\top}(I_2-\bar{I}_2)}{||I_1-\bar{I}_1||\,||I_2-\bar{I}_2||}, 
\end{equation}
where $\bar{I}_1$ and $\bar{I}_2$ are the means of the images.

\clearpage
\section{Partially coherent illumination} \label{sm:partially}

In this note we provide details about the definitions, computations and simulations associated with temporal and spatial coherence. We note that there exist multiple definitions in the literature for the coherence time and coherence length of an electromagnetic field. Here we follow the notations and the definitions used by Goodman~\citep{goodman2015statistical}. 

We start by defining the cross correlation function between two electromagnetic fields at two points in space lying on some plane and two time instances, $E(x_1,y_1,t+\tau)$ and $E(x_2,y_2,t)$, as
\begin{equation} \label{eq:cross_corr}
    \Gamma(x_1,y_1,x_2,y_2,\tau) = \langle E(x_1,y_1,t+\tau)E^*(x_2,y_2,t)\rangle,
\end{equation}
where $\langle\cdot\rangle$ denotes time average and $^*$ denote complex conjugation. The function $\Gamma(x_1,y_1,x_2,y_2,\tau)$ is also known as the mutual coherence function.
Using this definition we can define the complex degree of coherence, $\gamma(x_1,y_1,x_2,y_2,\tau)$, as
\begin{equation} \label{eq:complex_degree_of_coh}
    \gamma(x_1,y_1,x_2,y_2,\tau)=\frac{\Gamma(x_1,y_1,x_2,y_2,\tau)}{[I(x_1,y_1)I(x_2,y_2)]^{\frac{1}{2}}}\, ,
\end{equation}
where $I(x,y)=\langle |E(x,y,t)|^2\rangle$ denotes the intensity of the field at $(x,y)$. 
Equations~\eqref{eq:cross_corr} and~\eqref{eq:complex_degree_of_coh} are used for defining coherence time and coherence length, as we detail below. For further discussion we refer the readers to Goodman~\citep{goodman2015statistical} and to Mandel and Wolf~\citep{mandel1995optical}. 

\subsection{Temporal coherence}
\label{sm:partially-temporal}
Temporal coherence captures the decay of the correlation between an electromagnetic field at the same point in space and two different time instances, as a function of the time difference~$\tau$. Substituting $(x_1,y_1)=(x_2,y_2)\triangleq (x,y)$ in Eq.~\eqref{eq:cross_corr} and dropping the dependence on the spatial coordinates for simplicity, the autocorrelation for a time difference of $\tau$ is given by
\begin{equation} \label{eq:autocorr}
    \Gamma(\tau) = \langle E(x,y,t+\tau)E^*(x,y,t)\rangle.
\end{equation}

The function $\Gamma(\tau)$ is also known as the self coherence function. The complex degree of coherence in this case is
\begin{equation} \label{eq:t_complex_degree_of_coh}
    \gamma(\tau)=\frac{\Gamma(\tau)}{\Gamma(0)}.
\end{equation}

Following the definition proposed by Mandel~\citep{mandel1959fluctuations} and used by Goodman~\citep{goodman2015statistical}, the coherence time of the field, $\tau_c$, is defined as
\begin{equation} \label{eq:temporal_coh_def}
    \tau_c = \int_{-\infty}^{\infty}|\gamma(\tau)|^2 dt.
\end{equation}

As mentioned by Goodman~\citep{goodman2015statistical}, for a field with a Gaussian spectral density, as we used in our simulations, the integral in Eq.~\eqref{eq:temporal_coh_def} is given by

\begin{equation} \label{eq:temporal_coh}
    \tau_c = \frac{0.664}{\Delta \nu}, 
\end{equation}
where $\Delta \nu$ is the half power bandwidth in Hz of
the Gaussian spectrum, namely, the full-width-at-half-maximum (FWHM). The FWHM of a Gaussian distribution is given by 
\begin{equation} \label{eq:delta_nu}
   \Delta\nu=2\sqrt{2\ln 2}\,\sigma_\nu.
\end{equation}

In order to translate FWHM in Hz to wavelength bandwidth in meters, we use 
\begin{equation} \label{eq:delta_lambda}
   \Delta\lambda=\frac{c}{\nu_0-0.5\Delta\nu}-\frac{c}{\nu_0+0.5\Delta\nu}, 
\end{equation}
where $\nu_0$ is the central frequency of the Gaussian spectral distribution and $c$ is the speed of light.  
Using Eqs.~\eqref{eq:temporal_coh},~\eqref{eq:delta_lambda} we can analytically calculate the coherence time and the wavelength bandwidth for any given Gaussian spectral distribution.

In practice, we approximated the continuous spectrum by a discrete set of monochromatic sources with wavelengths picked uniformly at random from the effective support of the spectrum, each having an intensity proportional to the continuous spectrum at the corresponding wavelength. We chose the different $\sigma_\nu$ values in our experiments to obtain different numbers of discrete wavelengths, as we detail next.

We considered a Gaussian spectral distribution with central frequency $\nu_0$ and width $\sigma_\nu$,
\begin{equation} \label{eq:gaussian_nu}
    I(\nu) = \frac{1}{\sqrt{2\pi}\sigma_{\nu}} \exp\left\{-\frac{(\nu-\nu_0)^2}{2\sigma_{\nu}^2} \right\}. 
\end{equation}
Using the relation $\nu=\frac{c}{\lambda}$, we can express Eq.~\eqref{eq:gaussian_nu} as a function of $\lambda$ as

\begin{equation} \label{eq:gaussian_lambda}
    I(\lambda) = \frac{1}{\sqrt{2\pi}\sigma_{\nu}} \exp\left\{-\frac{(\frac{c}{\lambda_0})^2(\frac{\lambda_0-\lambda}{\lambda})^2}{2\sigma_{\nu}^2} \right\},
\end{equation}
where $\lambda_0=\frac{c}{\nu_0}$. 
We want to sample $I(\lambda)$, which is not a Gaussian, at $N_{\lambda}$ wavelengths $\{\lambda_i\}_{i=1}^{N_{\lambda}}$. 
For a given $\sigma_\nu$, 
we chose those wavelengths at random such that $\forall i\;I(\lambda_i)\geq\frac{1}{10}I(\lambda_0).$
We discarded wavelengths outside the visible spectrum of $400$ nm - $700$ nm. For each wavelength, $\lambda_{i}$, we took the intensity of the corresponding monochromatic source to be $I(\lambda_{i})$.

\begin{figure}
  \centering
   \includegraphics[width=\linewidth]{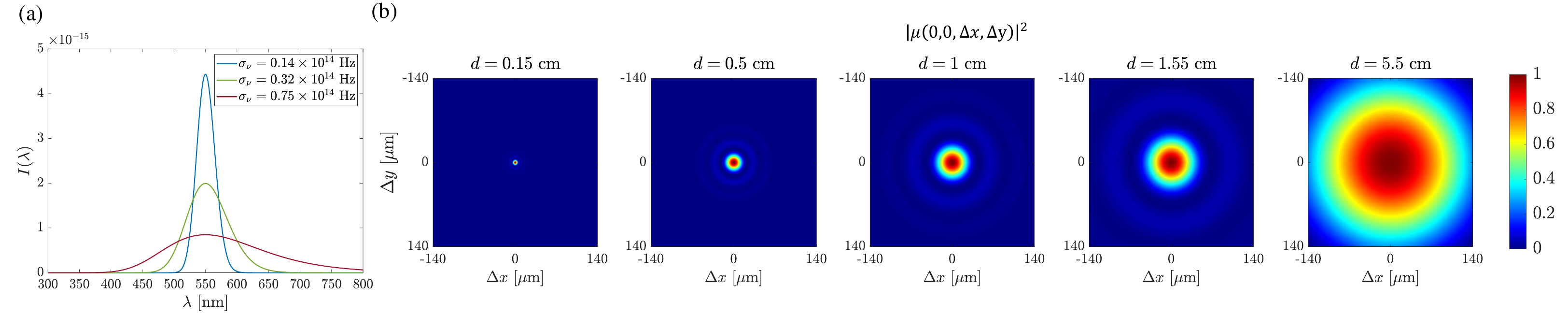}
   \caption{\textbf{Numerical calculation of temporal and spatial coherence.} (a) Gaussian spectral distribution of different width vs.~the wavelength, according to Eq.~\eqref{eq:gaussian_lambda}. 
   The curves shows a non-Gaussian shape, with notable tail to the right. (b) Plots of the integrand in Eq.~\eqref{eq:coh_area_x1y1}, for $(x_1,y_1)=(0,0)$, broadband source of $\Delta\lambda=75$ nm and for different $d$ values, mentioned in each plot title. As $d$ increases, $|\mu(x_1,y_1,\Delta x, \Delta y)|^2$ widens, indicating an increase in the coherence length $l_c$.}
   \label{fig:mu_i}
\end{figure}

We used Gaussian spectral distributions that were centered at $\nu_0=5.4\times10^{14}$~Hz, which translates to $\lambda_0=550$ nm. 
In order to work with settings corresponding to $N_{\lambda}=\{3,6,12\}$ wavelengths, we chose $\sigma_{\nu}=\{0.14, 0.32, 0.75\}\times10^{14}$ Hz. In addition to these multi-spectral distributions, we added a setting corresponding to a purely monochromatic source (namely with $\sigma_\nu=0$). From Eqs.~\eqref{eq:temporal_coh} and~\eqref{eq:delta_lambda}, the coherence times and corresponding bandwidths of the four resulting settings are $\tau_c = \{\infty, 19, 9, 4\}$ fs and $\Delta\lambda=\{0, 34, 75, 182\}$ nm, respectively. Figure~\ref{fig:mu_i}a displays the spectra $I(\lambda)$ of the three multi-spectral settings.  

\subsection{Spatial coherence}
\label{sm:partially-spatial}

The definition of spatial coherence relates to the correlation of the field between two different points in space at the same time instance. This definition follows by setting $\tau=0$ in Eqs.~\eqref{eq:cross_corr},\eqref{eq:complex_degree_of_coh}. 
Therefore, dropping the dependence on $\tau$, we are interested in the complex coherence factor,
\begin{equation} \label{eq:complex_coh_factor}
    \mu(x_1, y_1, x_2,y_2)=\gamma(x_1, y_1,x_2,y_2, 0).
\end{equation}

For \textit{quasi-monochromatic} light, the absolute value of the complex coherence factor depends only on the difference in coordinates $(\Delta x, \Delta y)=(x_2,y_2)-(x_1,y_1)$. Therefore, the coherence area of the field, $A_c$, is defined analogously to Eq.~\eqref{eq:temporal_coh_def} as \cite{goodman2015statistical}
\begin{equation} \label{eq:coh_area_def}
    A_c = \iint_{-\infty}^{\infty}|\mu(\Delta x, \Delta y)|^2 d\Delta x \,d\Delta y.
\end{equation}

Following the van Cittert–Zernike theorem \cite{zernike1938concept,van1934wahrscheinliche}, for a uniformly bright \textit{quasi-monochromatic} incoherent source of area $A_s$, located at a distance $d$ from the object plane, the coherence area $A_c$ on the object plane is given by
\begin{equation} \label{eq:coherence_area_sm}
    A_{c}=\frac{(\lambda d)^2}{A_{s}}.
\end{equation}

In order to work with coherence length, $l_c$, we can simply take the square root of the coherence area 
\begin{equation} \label{eq:coherence_length}
    l_{c}=\frac{\lambda d}{\sqrt{A_s}},
\end{equation}
where for a square source of side length $a$, like the one we used in our experiments, $\sqrt{A_s} = a$.

Equations~\eqref{eq:coh_area_def},~\eqref{eq:coherence_area_sm} are true only for a \textit{quasi-monochromatic} source of central wavelength $\lambda$.  
It turns out that for a broadband source, the coherence depends not only on $(\Delta x,\Delta y)$, but also on the coordinates themselves. Therefore, to associate a coherence area with a \textit{broadband} source, we first derive the coherence area corresponding to a fixed $(x_1,y_1)$ and a varying $(x_2,y_2)=(x_1,y_1)+(\Delta x, \Delta y)$. This coherence area depends on $(x_1,y_1)$. We then average this coherence area over points $(x_1,y_1)$ covering the relevant region in the object plane.

We start from the mutual intensity function, $J(x_1,y_1,x_2,y_2;\lambda)$, which follows by setting $\tau=0$ and using a single wavelength $\lambda$ in Eq.~\eqref{eq:cross_corr}. 
Following the Van Cittert-Zernike theorem, the mutual intensity is~\cite{goodman2015statistical}
\begin{equation}\label{eq:mutual_intensity}
    J(x_1,y_1,x_2,y_2;\lambda) =\frac{\kappa \exp[-j\psi(x_1,y_1,x_2,y_2,\lambda)]}{(\lambda d)^2}\iint_{-\infty}^{\infty}I(\xi,\eta)\exp\left[j\frac{2\pi}{\lambda d}(\Delta x \xi + \Delta y \eta)\right]d\xi d\eta.
\end{equation}

Here, $(\xi,\eta)$ are coordinates at the source plane, $\kappa$ is a constant with dimensions of length squared and $\psi(x_1,y_1,x_2,y_2,\lambda)$ is a phase factor given by
\begin{equation}\label{eq:psi}
    \psi(x_1,y_1,x_2,y_2,\lambda)=\frac{\pi}{\lambda d}\left[(x_2^2 + y_2^2) - (x_1^2 + y_1^2)\right].   
\end{equation}

The fields corresponding to different wavelengths are uncorrelated. Therefore, the mutual coherence function $\Gamma(x_1,y_1,x_2,y_2,0)$ of a broadband source is given by the sum (or integral) of the mutual intensities $J(x_1,y_1,x_2,y_2;\lambda)$ of the constituent wavelengths. Each mutual intensity function is summed with an intensity factor, $I(\lambda)$, matching to its wavelength, $\lambda$. As explained in Supplementary Note \ref{sm:partially-temporal}, $I(\lambda)$ is a sampled spectral Gaussian distribution.

Therefore, overall, we calculate the complex coherence factor of broadband illumination as 
\begin{equation}\label{eq:broadband_complex_coh_factor}
    \mu(x_1,y_1,x_2,y_2)=\frac{\sum_\lambda I(\lambda) \frac{\exp[-j\psi(x_1,y_1,x_2,y_2,\lambda)]}{\lambda^2}\iint_{-\infty}^{\infty}I(\xi,\eta)\exp\left[j\frac{2\pi}{\lambda d}(\Delta x \xi + \Delta y \eta)\right]d\xi d\eta}{\sum_\lambda I(\lambda) \frac{1}{\lambda^2}\iint_{-\infty}^{\infty}I(\xi,\eta)d\xi d\eta}.
\end{equation}

For a uniformly bright incoherent source of side $a$, the integral in the numerator is the 2D Fourier transform of a 2D rect$(\cdot,\cdot)$ function and the integral in the denominator is the area of the square source. Thus, Eq.~\eqref{eq:broadband_complex_coh_factor} can be simplified to 
\begin{equation}\label{eq:simplify_broadband_complex_coh_factor}
    \mu(x_1,y_1,x_2,y_2)=\frac{\sum_\lambda I(\lambda) \frac{\exp[-j\psi(x_1,y_1,x_2,y_2,\lambda)]}{\lambda^2}\text{sinc}(\frac{a\Delta x}{\lambda d},\frac{a\Delta y}{\lambda d})}{\sum_\lambda I(\lambda) \frac{1}{\lambda^2}}.
\end{equation}

In order to numerically calculate the coherence area, we define the point $(x_2,y_2)$ as $(x_1+\Delta x, y_1 + \Delta y)$, and numerically integrate  over $(\Delta x, \Delta y)$ to get an approximation of
\begin{equation} \label{eq:coh_area_x1y1}
    A_c(x_1,y_1) = \iint_{-\infty}^{\infty}|\mu(x_1,y_1,\Delta x, \Delta y)|^2 d\Delta x d\Delta y.
\end{equation}

We then average this coherence area over $(x_1,y_1)$. 

For numerically calculating Eq.~\eqref{eq:coh_area_x1y1} we used a grid of 1000 points between -140 $\mu$m to 140 $\mu$m for both $\Delta x$ and $\Delta y$. We then averaged the results over 100 different points of $(x_1,y_1)$, ranging between (0,0) and (140,140) $\mu$m. Plots of the integrand of Eq.~\eqref{eq:coh_area_x1y1} for $(x_1,y_1)=(0,0)$, broadband source of $\Delta\lambda=75$ nm and for different $d$ values are depicted in Fig.~\ref{fig:mu_i}b.

For all sources we used a set of five different values for $d$, getting different coherence lengths ranging from almost completely incoherent to coherent light. For all broadband sources we used $d=\{0.15, 0.5, 1, 1.55, 5.5\}$ cm, resulting in the coherence lengths of 
 \begin{equation}\label{eq:coher_lengths_as_funcs_of_bandwidth}
\begin{aligned}
    l_{c,\Delta\lambda=34\text{ nm}}&=\{6.7, 22.3, 44, 67.2, 204.7\}\;\mu\text{m},\\
    l_{c,\Delta\lambda=75\text{ nm}}&=\{6.6, 21.9, 43.4, 66.4,203.7\}\;\mu\text{m},\\
    l_{c,\Delta\lambda=182\text{ nm}}&=\{6.3, 20.8, 41.4, 63.6, 197.8\}\;\mu\text{m}.
\end{aligned}
\end{equation}

For the monochromatic source we used Eq.~\eqref{eq:coherence_length} and $d=\{0.14, 0.48, 0.94, 1.44, 4.36\}$ cm to achieve similar coherence lengths of 
\begin{equation}\label{eq:coher_lengths_for_monochrom}
    l_{c,\Delta\lambda=0}=\{6.5, 23, 43, 66, 200\}\;\mu\text{m}.
\end{equation}

\subsection{The effect of mismatch in temporal coherence between training time and test time}

We investigate the effect of mismatch in temporal coherence between training time and test time, similar to the investigation of spatial coherence depicted in Fig.~\ref{fig:s_test} in the main text. Unlike Fig.~\ref{fig:s_test}, where the spatial coherence varies and the temporal coherence stays fixed, here varying the temporal coherence causes slight covariation of the spatial coherence, as detailed above in Supplementary Note \ref{sm:partially-spatial}. The exact coherence lengths for each level of temporal coherence are provided in Eqs.~\eqref{eq:coher_lengths_as_funcs_of_bandwidth} and \eqref{eq:coher_lengths_for_monochrom}.  

\begin{figure}[t]
  \centering
   \includegraphics[width=\linewidth]{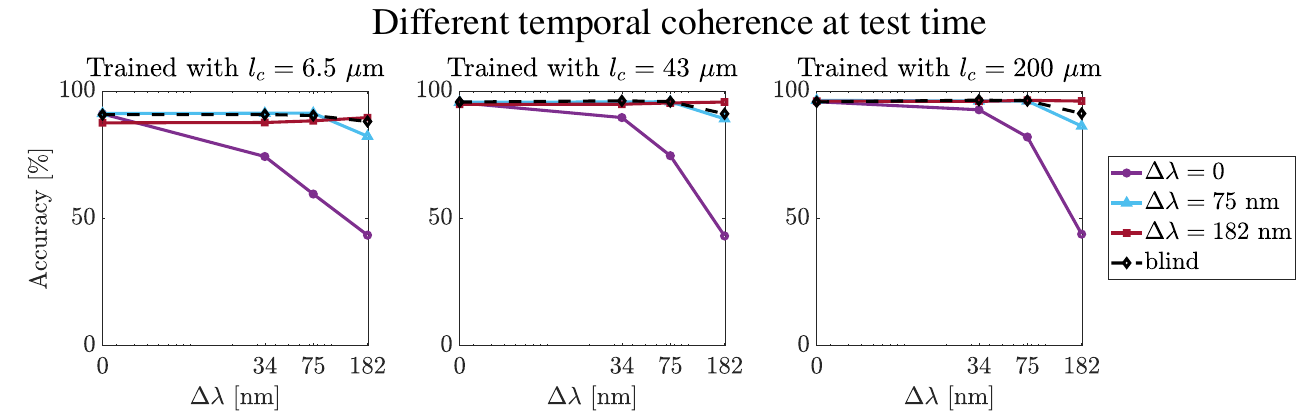}
   \caption{\textbf{The effect of mismatch in temporal coherence between training time and test time.} The network tends to show enhanced resilience to changes in temporal coherence settings. Yet, a substantial drop in performance is observed when testing a network that was trained for monochromatic light with broadband light spanning $182$ nm.}
   \label{fig:t_test}
\end{figure}

Figure~\ref{fig:t_test} illustrates the scenario of mismatched temporal coherence between training and test time. Each plot displays the results of three different diffractive networks trained with the same level of spatial coherence (specified in the title) but varying levels of temporal coherence (specified in the legend). While the performance peaks when there is zero mismatch, the drop in performance when deviating from this condition is not as significant as shown in Fig.~\ref{fig:s_test} when changing the spatial coherence conditions. Moreover, the decrease in performance for broadband sources trained with coherence lengths of approximately $l_c=\{43, 200\}\;\mu$m is nearly negligible. Therefore, the performance of these networks is similar to the performance of the blind networks. The network trained with incoherent illumination ($\Delta\lambda=182$ nm; red curve) even slightly surpasses the performance of the blind networks for some coherence conditions. Notably, the diffractive network trained with temporally coherent illumination ($\Delta\lambda=0$; purple curve) does experience a significant drop in performance when testing with different temporal coherence levels. For reference, each plot also depicts the accuracy of a coherence-blind network, which was trained for all temporal coherence conditions (black dashed line).

\subsection{Partial coherence effect on the FashionMNIST dataset}
In this section we provide additional results that illustrate the effect of spatial and temporal coherence on the performance of diffractive networks, as well as results of the effect of mismatch in spatial and temporal coherence between training time and test time. We again focus on two-layer linear diffractive networks for image classification with the same physical dimensions as mentioned in the main text, however, in this section we provide results for networks that were trained using the Fashion-MNIST dataset \citep{xiao2017fashion}. 

\begin{figure}
  \centering
   \includegraphics[width=\linewidth]{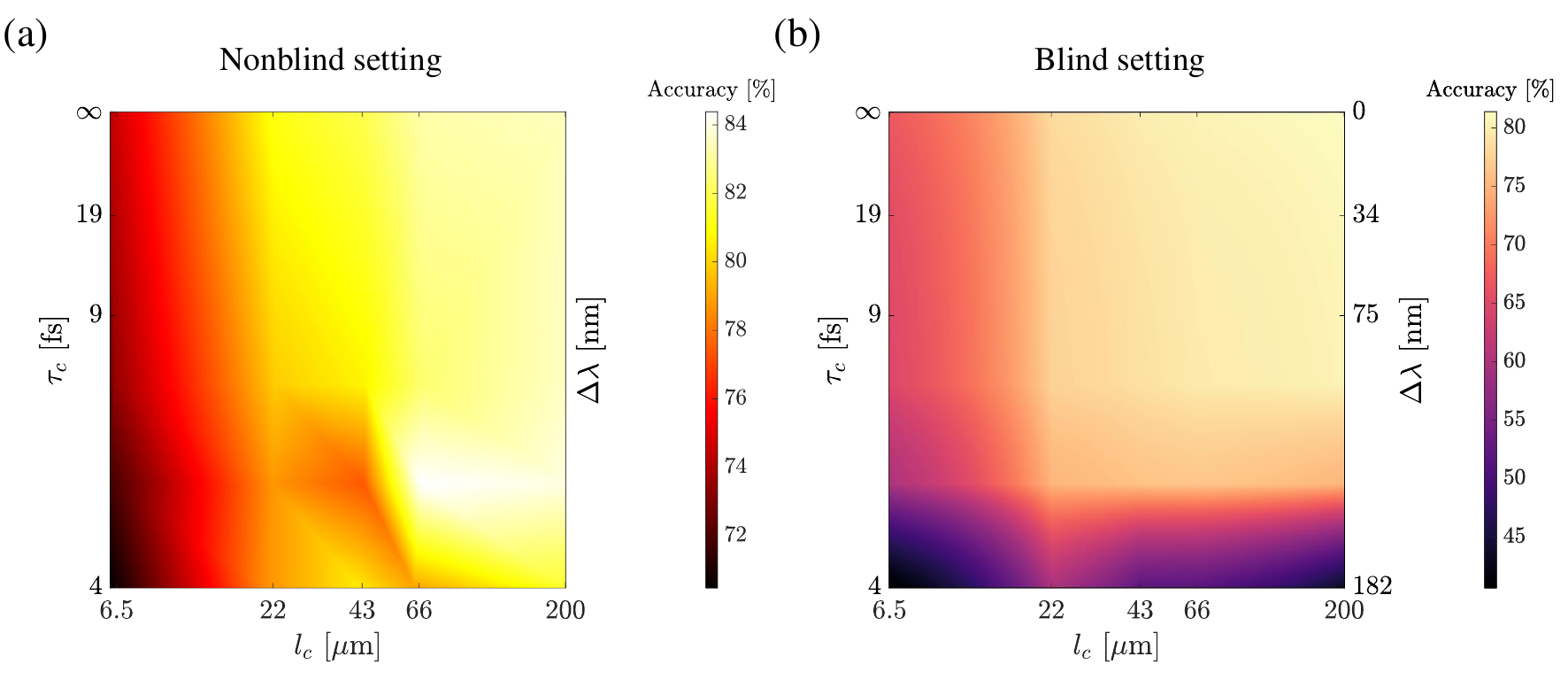}
   \caption{\textbf{Effect of spatial and temporal coherence on the performance of diffractive networks trained on the FashionMNIST dataset.} The figure depicts the accuracy rates for different levels of spatial and temporal coherence, both for the nonblind (a) and for the blind (b) settings. The horizontal axis measures spatial coherence via coherence length. The vertical axis measures temporal coherence and has two scales, one for coherence time (left) and another for the associated wavelength bandwidth around the central wavelength of 550 nm (right). Both axes are on a logarithmic scale. The different colors on the graph indicate the resulting accuracy.}
   \label{fig:fashion_partial}
\end{figure}

Figure~\ref{fig:fashion_partial} illustrates the effect of spatial and temporal coherence on the performance of diffractive networks, similarly to Fig.~\ref{fig:partial}, but for the Fashion-MNIST dataset. We numerically examine both the nonblind and the blind settings. 
In the nonblind case shown in Fig.~\ref{fig:fashion_partial}a, the classification accuracy ranges from 70.3\% to 84.4\%, and, as illustrated in Fig.~\ref{fig:partial}, depends more strongly on the degree of spatial coherence than on the degree of temporal coherence. For almost all degrees of temporal coherence, the results improve as the spatial coherence grows larger.   

Figure~\ref{fig:fashion_partial}b shows the results achieved by a single coherence-blind network, which was trained on all coherence settings simultaneously.
The accuracy rates range from 40.5\% to 81.4\%, and as in the nonblind setting, they depend more strongly on the degree of spatial coherence than on the degree of temporal coherence. As in Fig.~\ref{fig:partial}, the extreme range of coherence levels we consider is expected to exceed that encountered in any realistic application. Therefore, these results can be considered as a lower bound on the accuracy that can be expected in practical settings. In particular, the lowest accuracy is obtained when testing the blind network on temporally incoherent illumination with different spatial coherence conditions. 
Excluding these results, the accuracy rates range from 60.3\% to 81.4\%, which is still a significant drop w.r.t~the nonblind setting, but not as severe.   

\begin{figure}
  \centering
   \includegraphics[width=\linewidth]{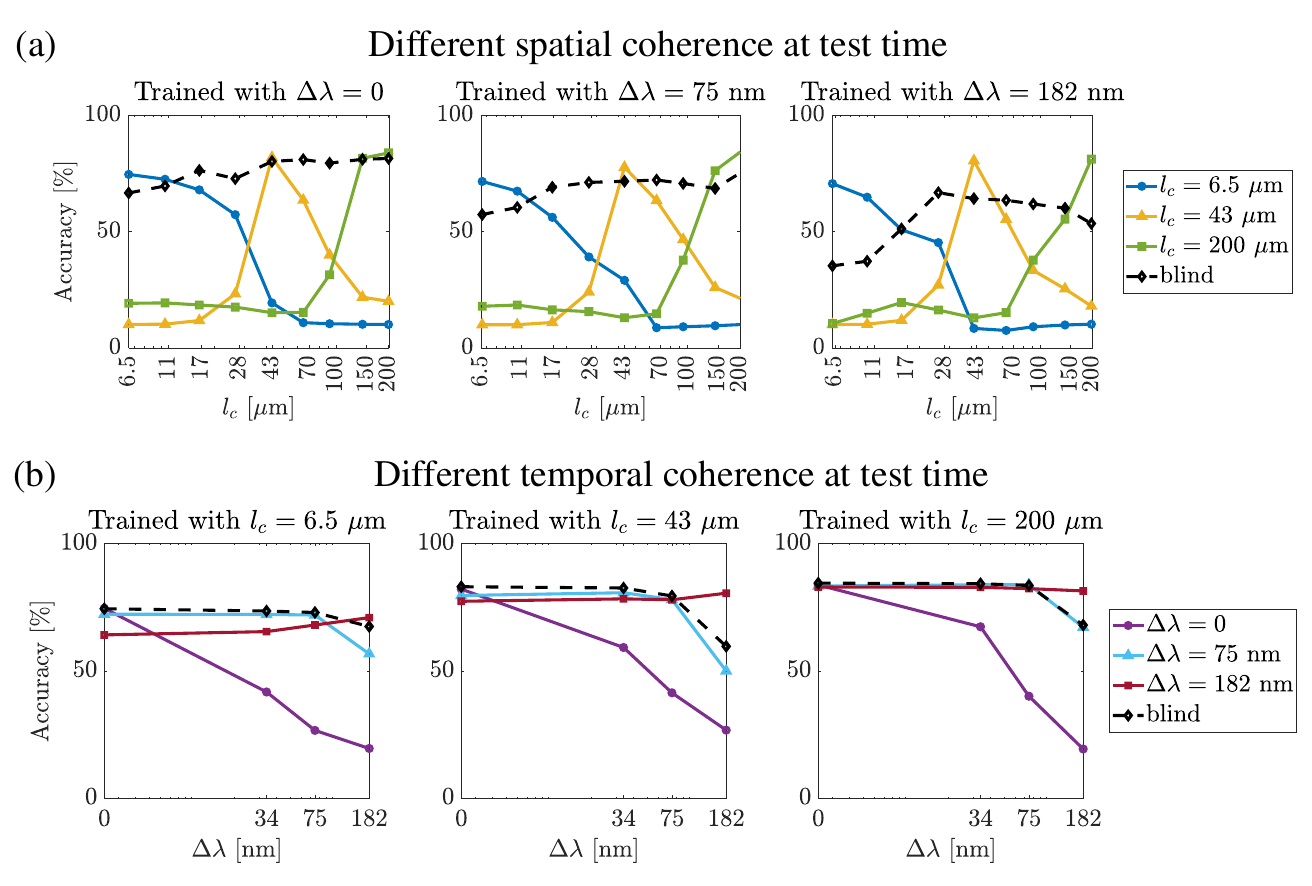}
   \caption{\textbf{The effect of mismatch in spatial and temporal coherence between training time and test time on the FashionMNIST dataset.} (a) Each plot displays the results of three different diffractive networks, all trained with the same level of temporal coherence (specified in the title) and with different levels of spatial coherence (specified in the legend). For reference, each plot also depicts a spatial coherence blind network as a black dashed line. This spatial coherence blind network was trained on the same level of temporal coherence as the nonblind networks. (b) The same as in (a), but here the spatial coherence is the same for each plot, which depicts the accuracy of each network as a function of the level of temporal coherence.}
   \label{fig:fashion_s_t_test}
\end{figure}

Figure \ref{fig:fashion_s_t_test} illustrates the effect of mismatch in coherence conditions between training time and test time, for both spatial (Fig.~\ref{fig:fashion_s_t_test}a) and temporal (Fig.~\ref{fig:fashion_s_t_test}b) coherence. The effect seen in Fig.~\ref{fig:fashion_s_t_test} is similar to the one seen in Fig.~\ref{fig:s_test} and Fig.~\ref{fig:t_test}. However, as can be seen in the rightmost plot in Fig.~\ref{fig:fashion_s_t_test}a, a network trained to accommodate different spatial coherence levels for temporally incoherent illumination (black dashed curve) is not as robust to the level of spatial coherence as its counterparts which were trained with temporally coherent (left plot) and partially coherent (middle plot) illumination. Nevertheless, this blind network still outperforms the nonblind networks when the mismatch in spatial coherence is large. As explained above, this network was trained on a large range of spatial coherence conditions, which is not expected in a real scenario, and hence its performance can be considered a lower bound on the performance expected in real-world settings. In addition, similarly to Fig.~\ref{fig:t_test}, the performance of the networks trained under partial temporal coherence, and specifically the network trained with temporally incoherent illumination ($\Delta\lambda=182$ nm; red curve), are almost as good as those of the blind network.

\clearpage

\section{The incoherent approximation}
\label{sm:limits}

In this note we discuss when an optical system can be considered as operating under fully incoherent illumination. We start by discussing imaging systems and then extend the discussion to a more general class of optical systems, which include diffractive networks. As in \ref{sm:partially}, we follow the definitions and derivations used by Goodman~\citep{goodman2015statistical}. 

\subsection{The intensity at the output of an optical system}
Consider a space-invariant optical system.
This assumption is often valid~\citep{goodman2015statistical} given that one properly scales (and optionally inverts) the axes at the output plane. Let us denote by $J_o$ the mutual intensity function (as defined in Supplementary Note \ref{sm:partially-spatial}) just before the object plane, by $t_o$ the object amplitude transmittance, and by $K$ the amplitude point-spread function of the system from the object to the output plane. Then the intensity at the output plane can be written as~\citep{goodman2015statistical} 
\begin{align}
\label{eq:space-inv-intensity}
I_{\text{out}}(x',y')=\iiiint_{-\infty}^{\infty}&K(x'-x,y'-y)K^*(x'-x-\Delta x,y'-y -\Delta y) \nonumber \\
& \times t_o(x,y)t_{o}^{*}(x+\Delta x,y + \Delta y) J_{o}(\Delta x, \Delta y) d x d y d\Delta x d\Delta y. 
\end{align}
As shown by Goodman~\citep{goodman2015statistical}, this relation can also be written in the Fourier domain as
\begin{align}
\label{eq:space-inv-intensity-freq}
\mathcal{I}_{\text{out}}(u,v)=\iiiint_{-\infty}^{\infty}&\mathcal{K}(s - s',t-t')\mathcal{K}^*(s-s'-u,t-t'-v) \nonumber \\
&\times \mathcal{T}_o(s,t)\mathcal{T}_{o}^{*}(s - u,t - v) \mathcal{J}_{o}(s',t') d s d t ds' dt',
\end{align}
where $\mathcal{I}_{\text{out}}$, $\mathcal{K}$, $\mathcal{T}_{o}$ and $\mathcal{J}_{o}$ denote the Fourier transforms of $I_{\text{out}}$, $K$, $t_{o}$ and $J_{o}$, respectively.%

When the light at the object plane is incoherent, $J_o(\Delta x,\Delta y)=\kappa I_o \delta(\Delta x,\Delta y)$, so that $\mathcal{J}_o$ is the constant function 
\begin{equation}
\mathcal{J}_o(s',t')=\kappa I_o.
\end{equation}
Therefore, the question of whether a system can be considered as operating under incoherent illumination, translates into the question of whether the coherence function $\mathcal{J}_o$ at the object plane can be replaced by a constant function without changing the result of the integral in Eq.~\eqref{eq:space-inv-intensity-freq}. For this to happen, it suffices for $\mathcal{J}_o$ to be constant only over the domain where the other terms of the integrand do not vanish. 
If the object is illuminated by a planar, uniformly bright, incoherent source with pupil function $P_i$, then we have from the Van Cittert-Zernike theorem that $\mathcal{J}_o$ can be written as 
\begin{equation}\label{eq:fourier_J}
    \mathcal{J}_o(s',t') = \kappa I_o |P_i(-\lambda d s',-\lambda d t')|^2,
\end{equation}
where, $I_o$ is a constant intensity and $d$ is the distance between the source plane and the object plane. Let us assume for simplicity that the pupil $P_i$ is a disk of diameter $a$. In this case, the system is in the incoherent regime when the radius $a / 2$ is large enough to contain the supports of all other terms in the integrand in Eq.~\eqref{eq:space-inv-intensity-freq}. We next discuss when this happens for imaging systems, and then extend the discussion to systems with a random phase masks.

\subsection{The incoherence regime for an imaging system}
We begin by discussing the case of an optical system that performs imaging using a single lens. For simplicity, we focus on the case of unit magnification (otherwise the axes at the output plane need to be properly scaled). The amplitude point-spread function, $K$, of such a system can be written as 
\begin{equation}\label{eq:ASF_spa_inv}
K(x',y')=\frac{1}{(\lambda d_e)^2} \iint_{-\infty}^{\infty} P(\xi,\eta) \exp\left\{-j\frac{2\pi}{\lambda} \left(\frac{\xi x'}{d_e} + \frac{\eta y'}{d_e}\right)\right\} d\xi d\eta,
\end{equation}
where $(\xi,\eta)$ are the coordinates at the plane of the lens, and $(x',y')$ are the coordinates at the output plane. Here, $d_e$ denotes the distance between the lens and the output plane, and $P(\xi,\eta)$ is the pupil function of the lens, defined as
\begin{equation}    
 P(\xi,\eta) = \begin{cases}
     1 & \text{ if } (\xi,\eta) \text{ is inside the aperture},\\
     0 & \text{otherwise}.
 \end{cases}
\end{equation}
Following Goodman~\cite{goodman2015statistical}, we take $P$ to be a circular aperture. A similar analysis can be done for a rectangular aperture. 
It can be seen from Eq.~\eqref{eq:ASF_spa_inv} that $K$ is the Fourier transform of the pupil function $P$. Therefore, $\mathcal{K}$, the Fourier transform of $K$, is a scaled version of the pupil function, 
\begin{equation}\label{eq:fourier_K}
    \mathcal{K}(s,t) = P(\lambda d_e s,\lambda d_e t).
\end{equation}

\begin{figure}[t]
    \centering
    \includegraphics[width=0.5\linewidth]{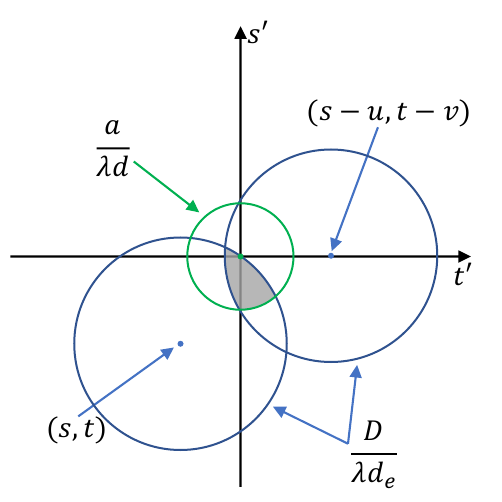}
    \caption{\textbf{Visual explanation of Eq.~\eqref{eq:space-inv-intensity-freq}}. The blue circles, each of diameter $\frac{D}{\lambda d_e}$ (see Eq.~\eqref{eq:fourier_K}) depict the displaced point-spread functions. The green circle, of diameter $\frac{a}{\lambda d}$, is the result of Eq.~\eqref{eq:fourier_J}. The circles are drawn in the frequency domain, hence their diameter dimensions is one over length. The area of overlap, marked in gray, is the area computed by Eq.~\eqref{eq:space-inv-intensity-freq}.}
    \label{fig:overlap}
\end{figure}

Let us examine the integrals with respect to $s',t'$ in Eq.~\eqref{eq:space-inv-intensity-freq}. For every $s,t,u,v$, we have a product of two shifted discs, corresponding to the terms associated with $\mathcal{K}$ (as we have just seen from Eq.~\eqref{eq:fourier_K} that $\mathcal{K}$ is a circ function). These two discs are indicated by the blue circles in Fig.~\ref{fig:overlap}. Furthermore, the term with $\mathcal{J}_o$ is also a circ function (see Eq.~\eqref{eq:fourier_J}), which is shown as a green circle in Fig.~\ref{fig:overlap}. Therefore, to be in the incoherent regime,  the green circle, representing $\mathcal{J}_o$, must grow until it completely covers the intersection of the two blue circles. The largest radius required of the source illumination occurs when the two blue circles completely overlap (\textit{i.e.} when $(u,v)=(0,0)$) and their joint center $(s,t)$ is as far as possible from the origin. In this case, the radius of the green circle must be larger than the distance of the center of the blue circles from the origin, $\sqrt{s^2+t^2}$, plus the radius of the blue circles, $D/(2\lambda d_e)$. Namely,
\begin{equation}
     \frac{a/2}{\lambda d} = \frac{D}{2\lambda d_e} + \max \{\sqrt{s^2+t^2}\}.
\end{equation}

To determine the maximal value of $\sqrt{s^2+t^2}$ we need to examine the Fourier transform of the object's transmittance, $\mathcal{T}_o$. Note that for $(u,v)=(0,0)$, the terms associated with $\mathcal{T}_o$ in Eq.~\eqref{eq:space-inv-intensity-freq} reduce to $|\mathcal{T}_o(s,t)|^2$. Assuming that $\mathcal{T}_o$ is supported on a disk of radius $f^{\text{obj}}_{\max}$, the term $|\mathcal{T}_o(s,t)|^2$ vanishes when $\sqrt{s^2+t^2} > f^{\text{obj}}_{\max}$, so that $\max \{\sqrt{s^2+t^2}\} = f^{\text{obj}}_{\max}$. Therefore, a system behaves as if the object emits incoherent light whenever 
\begin{equation} \label{eq:incoh-lim-freq}
     \frac{a/2}{\lambda d} \ge \frac{D}{2\lambda d_e} + f^{\text{obj}}_{\max},
\end{equation}
or, equivalently, whenever
\begin{equation}
    \frac{\lambda d}{a/2} \le \frac{1}{\frac{D}{2\lambda d_e} + f^{\text{obj}}_{\max}}.
\end{equation}
We note that $\frac{\lambda d}{\sqrt{\pi}a/2}$ is equal to the coherence length $l_c$ on the object plane (see Supplementary Note \ref{sm:partially-spatial}). In addition, as we analyze a unit-magnitude imaging system, the distance between the lens and the output plane is equal to the distance between the object and the lens, namely $d_i=d_e$. Furthermore, $\frac{D}{2\lambda d_i}$ is precisely the maximal spatial frequency preserved by the system, which we denote by $f^{\text{sys}}_{\max}$. Using these observations, the condition for a  system to be in the incoherent regime can be compactly written as 
\begin{equation} \label{eq:incoh-lim}
   l_c \le \frac{1}{\sqrt{\pi}} \cdot \frac{1}{ f^{\text{sys}}_{\max} + f^{\text{obj}}_{\max}}.
\end{equation}

\subsection{Optical systems containing a random phase mask}

We next discuss a more general class of optical systems, which instead of a lens, contain a phase mask $\phi(\xi,\eta)$ with random phases. As we will see, the same condition characterizes when such systems are in the incoherent regime.

Specifically, we consider a diffractive element with transfer function $t_d(\xi,\eta) = P(\xi,\eta)\exp\{-j\frac{\pi}{\lambda d_e}\phi(\xi,\eta)\}$, where $\phi(\xi,\eta)$ is a random phase mask. To cancel out the quadratic phase elements in the resulting amplitude point-spread function, let us add and subtract a quadratic phase term, 
\begin{equation}
    t_d(\xi,\eta) = P(\xi,\eta)\exp\left\{-j\frac{\pi}{\lambda d_e}(\xi^2+\eta^2-\xi^2-\eta^2+\phi(\xi,\eta))\right\}.
\end{equation}
This shows that the amplitude point-spread function $K$, which was given in Eq.~\eqref{eq:ASF_spa_inv} for imaging systems, now becomes
\begin{align}\label{eq:ASF_spa_inv_rand}
K(x',y')=\frac{1}{(\lambda d_e)^2} \iint_{-\infty}^{\infty} P(\xi,\eta)\exp\left\{-j\frac{\pi}{\lambda d_e}(\phi(\xi,\eta)+\xi^2+\eta^2)\right\} \nonumber\\
 \times \exp\left\{-j\frac{2\pi}{\lambda} \left(\frac{\xi x'}{d_e} + \frac{\eta y'}{d_e}\right)\right\} d\xi d\eta.
\end{align}
Therefore, in this setting $K$ is the Fourier transform of the pupil function multiplied by a random and quadratic phase. Therefore, $\mathcal{K}$ the Fourier transform of $K$, is a scaled version of the pupil function multiplied by a random and quadratic phase. Substituting this into Eq.~\eqref{eq:space-inv-intensity-freq}, gives 
\begin{align}
&\mathcal{I}_{\text{out}}(u,v)=\iiiint_{-\infty}^{\infty} \mathcal{T}_o(s,t)\mathcal{T}_{o}^{*}(s-u,t - v) \mathcal{J}_{o}(s',t') \nonumber\\
& \times P^*(s - s' - u,t-t'-v)\exp\left\{j\frac{\pi}{\lambda d_e}\left(\phi(s - s'-u,t-t'-v)+(s-s'-u)^2+(t-t'-v)^2\right)\right\} \nonumber\\
& \times P(s - s',t-t')\exp\left\{-j\frac{\pi}{\lambda d_e}\left(\phi(s - s',t-t')+(s-s')^2+(t-t')^2\right)\right\} d s d t ds' dt'. 
\end{align}
Let us take the expectation of this expression with respect to the random phases. Due to the linearity of the integral and the expectation, we can switch their orders. Furthermore, since $\phi(\xi,\eta)$ is the only random element within the integrand, we need only take the expectation of the term $\exp\{j\frac{\pi}{\lambda d_e}(\phi(s - s'-u,t-t'-v)\}\exp\{-j\frac{\pi}{\lambda d_e}(\phi(s - s',t-t')\}$. We assume that by construction, the phase map is drawn from a distribution such that this expectation depends only on the shift between the terms, $(u,v)$, and denote this expectation by $\Gamma_{\phi}(u,v)$. Note that since the autocorrelation function $\Gamma_{\phi}(u,v)$ does not depend on $s,t,s',t'$, it can be pulled out of the integral, which simplifies the expression to
\begin{align}\label{eq:intensity-freq-auto}
\mathcal{I}_{\text{out}}&(u,v)=\Gamma_{\phi}(u,v) \iiiint_{-\infty}^{\infty} \mathcal{T}_o(s,t)\mathcal{T}_{o}^{*}(s-u,t - v) \mathcal{J}_{o}(s',t') \nonumber \\
& \times P^*(s - s' - u,t-t'-v) P(s - s',t-t')\nonumber \\
& \times \exp\left\{-j\frac{\pi}{\lambda d_e}\left((s-s')^2+(t-t')^2-(s-s'-u)^2-(t-t'-v)^2\right)\right\} d s d t ds' dt'. 
\end{align}
Note that the integral in this expression is identical to that appearing in Eq.~\eqref{eq:space-inv-intensity-freq}. Therefore, we can follow the precise same arguments to obtain the condition for being in the incoherent regime. In particular, the maximal overlap between the blue circles in Fig.~\ref{fig:overlap} is maximal when $(u,v)=(0,0)$, and by definition, $\Gamma_{\phi}(0,0)>0$, so that this is a valid point. Therefore, the incoherent approximation conditions do not change for optical systems containing random phase patterns.

\clearpage

\section{Effect of different design choices on incoherent diffractive networks}
\label{sm:incoherent light}

In this note we further investigate the performance of diffractive networks trained under incoherent illumination. Specifically, we investigate how the performance depends on the input object size and on the number of diffractive layers within the network. In addition, we also investigate the effect of the number of wavelengths used during training on the performance of the diffractive network. 

In all the numerical experiments in this note we trained diffractive networks on the MNIST dataset with the same hyperparameters mentioned in the main text, unless specified otherwise. 

\subsection{Varying the input object size}

We start by examining linear and nonlinear networks with a single diffractive layer. We trained four different networks with object dimensions $(W_o, H_o) = (280,280)$, $(500,500)$, $(600,600)$, $(700,700)$  $\mu$m. These objects, if needed, were first upsampled from their original dimensions using nearest neighbor interpolation. 

\begin{figure}[t]
  \centering
   \includegraphics[width=\linewidth]{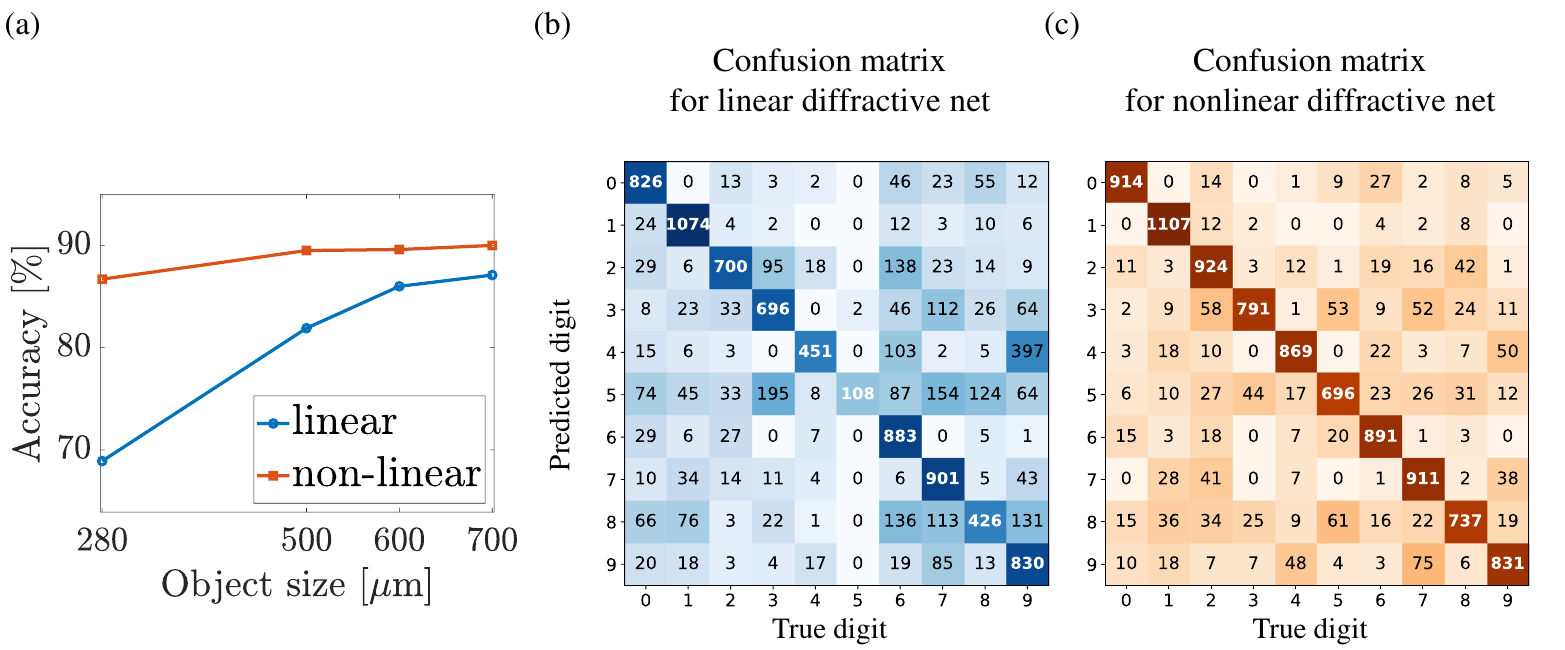}
   \caption{\textbf{Effect of training with different number.} (a) Comparison between linear and nonlinear single-layer diffractive networks, as a function of the object dimensions. (b) Confusion matrix for linear diffractive network with object dimensions of $(W_o, H_o) = (280,280)$ $\mu$m. Here, the model's predictions are often not accurate, especially for the digits 5 and 8. (c) Confusion matrix for a nonlinear diffractive network with object dimensions of $(W_o, H_o) = (280,280)$ $\mu$m. Here, the diffractive network achieves good performance.}
   \label{fig:varying_object}
\end{figure}

Comparison between the accuracy achieved by both linear and nonlinear networks for different object dimensions is provided in Fig.~\ref{fig:varying_object}a. The accuracy increases with object size, from $68.9\%$ to $87.1\%$ for the linear setting, and from $86.7\%$, to $90\%$ for the nonlinear setting. Computation cost and running time also grow larger with the increased dimensions.  

For the object dimensions of $(W_o, H_o) = (280,280)$ $\mu$m we also depict the classification results of the linear and nonlinear network in the structure of a confusion matrix (Fig.~\ref{fig:varying_object}b,c). The vertical and horizontal axes align with the true and predicted labels, respectively. Each cell in a confusion matrix represents a specific type of prediction made by the diffractive network. The value in each cell indicates the count of instances that fall into each category. The higher the value in each cell, the deeper its color. The diagonal of the matrix is where the diffractive network's prediction and the true labels coincide, thus the deeper the color of the diagonal, the better the overall performance.    

\subsection{Varying the number of diffractive layers} 

We further investigated the effect of using deeper linear diffractive networks on the classification accuracy. Adding more diffractive layers entails incorporating additional learnable parameters for optimization and involves extra free space propagation between consecutive layers. 
We first trained diffractive networks with two diffractive layers for object dimensions of $(W_o, H_o) = (280,280),(500,500),(600,600),(700,700)$ $\mu$m. These networks achieved  between $87.5\%$ and $90.2\%$ test accuracy. 

We further increased the number of diffractive layers to 3, 4 and 5. Due to computational constraints, for these deeper networks, we studied only objects of size $(W_o, H_o) = (280,280)$ $\mu$m. The achieved test accuracy was $88.9\%, 87.6\%, 80.6\%$, for 3, 4, and 5 layers respectively. While increasing the number of layers from 2 to 3 increases also the accuracy, increasing it further deteriorates the results. 

This deterioration can be attributed to vanishing gradients in the diffractive networks. The phenomenon of vanishing gradients, which refers to the situation in which the gradients in neural networks are very small, effectively prevents the weights from updating during the optimization \cite{hochreiter2001gradient}. Vanishing gradients can also occur in diffractive networks, especially for deeper networks \cite{mengu2019analysis}. This deterioration can also be attributed to the hyperparameters we used for training the deeper networks, as those hyperparameters were tuned for a single diffractive layer network. 

Those issues can potentially be mitigated using different optimization schemes and further hyperparameter tuning. We leave this for future work.

\subsection{The effect of the number of wavelengths used for training}

We also investigated the effect of the number of wavelengths used during training on the performance of the trained diffractive networks. Specifically, we trained diffractive networks with two layers under incoherent illumination, with object dimensions of $(W_o, H_o) = (280,280)$ $\mu$m. As opposed to the Gaussian spectrum used in our other experiments, here we trained the networks with 3, 6 and 12 wavelengths, uniformly sampled from a \emph{flat spectrum} between $400-700$ nm. The accuracy achieved by those networks when testing on a spectrum with the same wavelengths as in training were 86.5\%, 87.5\% and 88.2\%, respectively. This illustrates that increasing the number of wavelengths during training improves the results, although it also increases the training time. In particular, doubling the number of wavelengths causes the training time to be approximately twice as slow.

\begin{figure}[t]
  \centering
   \includegraphics[width=0.7\linewidth]{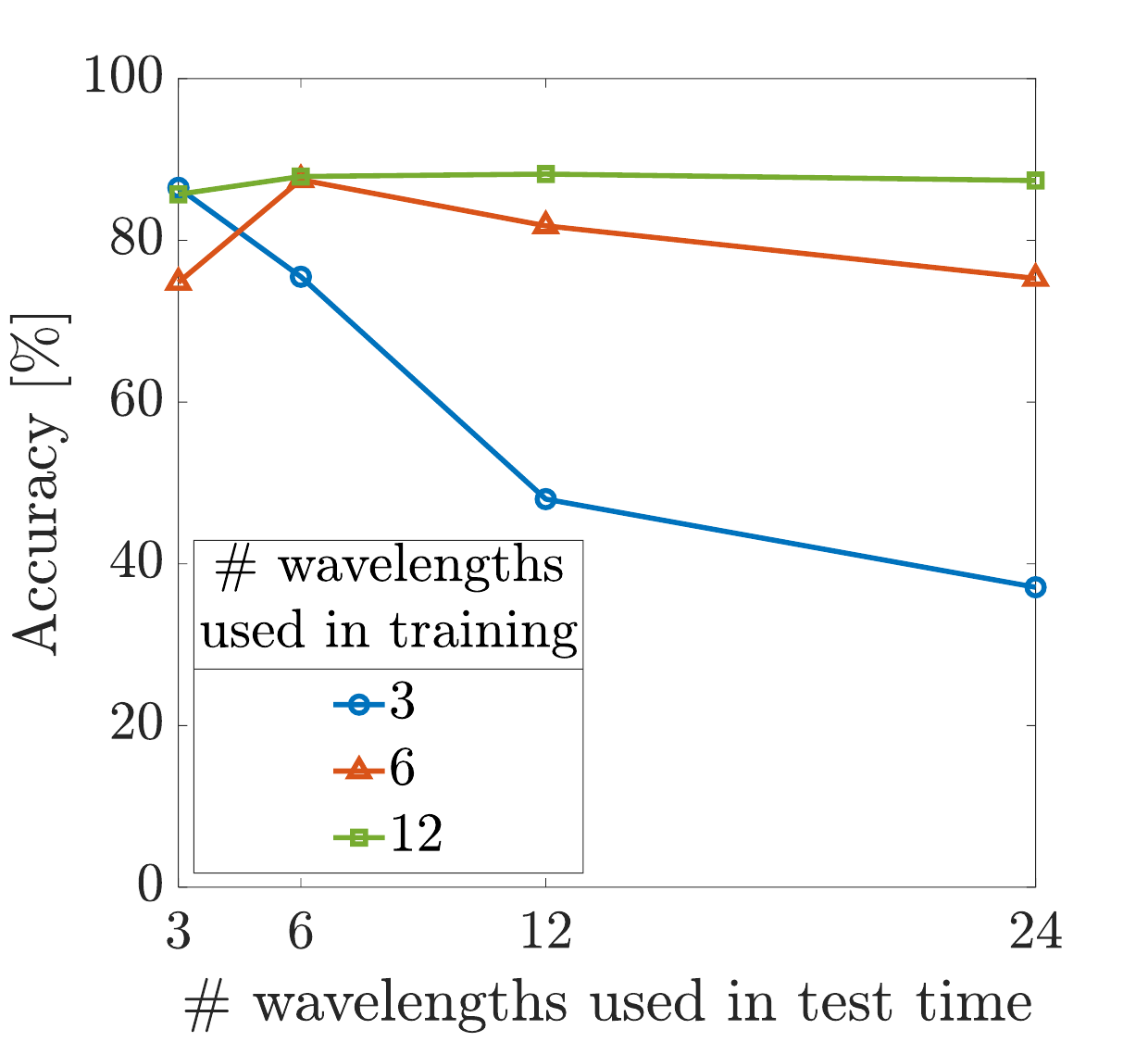}
   \caption{\textbf{The effect of mismatch in temporal coherence between training time and test time for spatially incoherent illumination with a flat spectrum.} Each curve displays the accuracy of a diffractive network trained with uniformly sampled wavelengths from a flat spectrum. The number of wavelengths used for training is mentioned in the legend. The horizontal axis indicates the number of wavelengths used at test time.}
   \label{fig:wavelengths_num}
\end{figure}

Using more wavelengths during training also improved the resilience of the diffractive networks for mismatch in the number of wavelengths between training time and test time. For each network, we evaluated its results when using different numbers of wavelengths at test time. The wavelengths used at test time were also uniformly sampled from a flat spectrum between $400-700$ nm. The results are shown in Fig.~\ref{fig:wavelengths_num}. 

The accuracy of the network that was trained with 3 wavelengths significantly decreases when there is a mismatch in the number of wavelengths between training time and test time. Specifically, for inference with 24 wavelengths this network suffers from a 49.4\% drop in its performance. The network that was trained with 6 wavelengths shows more resilience to a mismatch in the spectrum, but still suffers from up to a 12.7\% drop in its performance. The network that was trained with 12 wavelengths suffers the least from a mismatch and its biggest drop in performance is only 2.5\%.    

\clearpage
\section{Numerical fabrication effect}
\label{sm:quant}
In order to fabricate the learned diffractive layers, the learned phase, $\phi(x,y)$, needs to be translated into height maps. This translation uses the known relation $h(x,y) = \frac{\lambda}{2\pi n}\phi(x,y)$, where $n$ is the refractive index of the material used for fabrication~\cite{Lin2018}. 
Current processing technology supports discrete heights, thus requiring to first quantize the phase into a discrete set of values~\cite{chen2021diffractive}.

To study the effect of the quantization, we quantized the learned phases into $q$ levels, evenly spaced between $[0,2\pi]$, where $q=\{2, 4, 8, 16, 32\}$. We evaluated the edge cases A,B,C,D of Fig.~\ref{fig:partial} in both the blind and the nonblind setting, after this numerical fabrication. 

\begin{figure}[h]
  \centering
   \includegraphics[width=\linewidth]{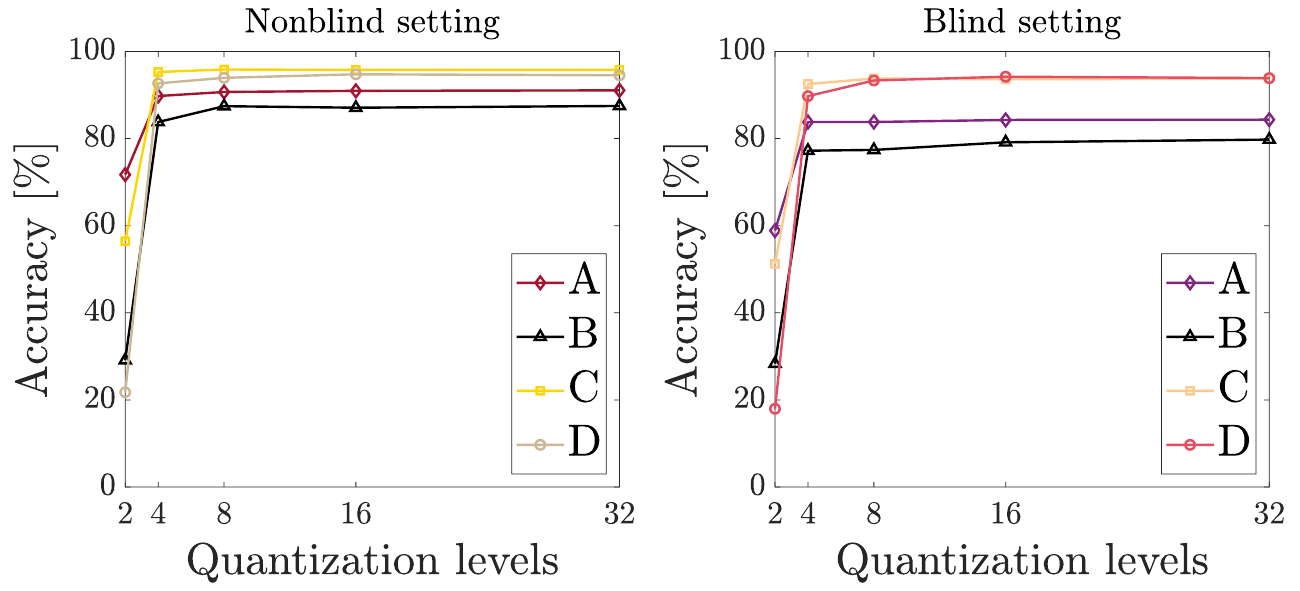}
   \caption{\textbf{Numerical experiment for studying fabrication effects.} The plots depict the accuracy of four networks in the nonblind and blind coherence settings, as a function of the quantization level of the phase mask. When using 4 or more quantization levels, the accuracy is almost the same as with the full-precision phase values.}
   \label{fig:quant}
\end{figure}

As can be seen in Fig.~\ref{fig:quant} there is a substantial decrease in performance for 2 quantization levels, for both the blind and non blind settings. There is a relatively small drop in performance for 4 quantization levels and almost no drop in performance for 8 or more quantization levels, again for both settings. 

\clearpage
\section{Comparison to modal expansion}
\label{sm:modal expansion}

Modal expansion\cite{wolf1982new} is a common method for analyzing the free space propagation of incoherent electromagnetic fields. This method was used by Rhaman et al.\citep{rahman2023universal} in a recent work for training a spatially incoherent diffractive network. Modal expansion involves taking an input electromagnetic field $E$ and multiplying it by some random phase, $\exp\{j\phi_r(x,y)\}$, where, $\phi_r(x,y)\sim \mathcal{U}([0,2\pi])$ independently for each $(x,y)$. The field is coherently propagated through a diffractive network and the propagation is repeated for $N_{d}$ random draws, each time with different random phase $\phi_r(x,y)$. The resulting intensities are averaged over the different draws to obtain an approximation of the intensity that is obtained for an incoherent source. As the number of iterations $N_{d}$ increases, the approximation becomes more accurate, and in the limit where $N_{d} \rightarrow \infty$, the approximation is exact. 

\begin{figure}
  \centering
   \includegraphics[width=\linewidth]{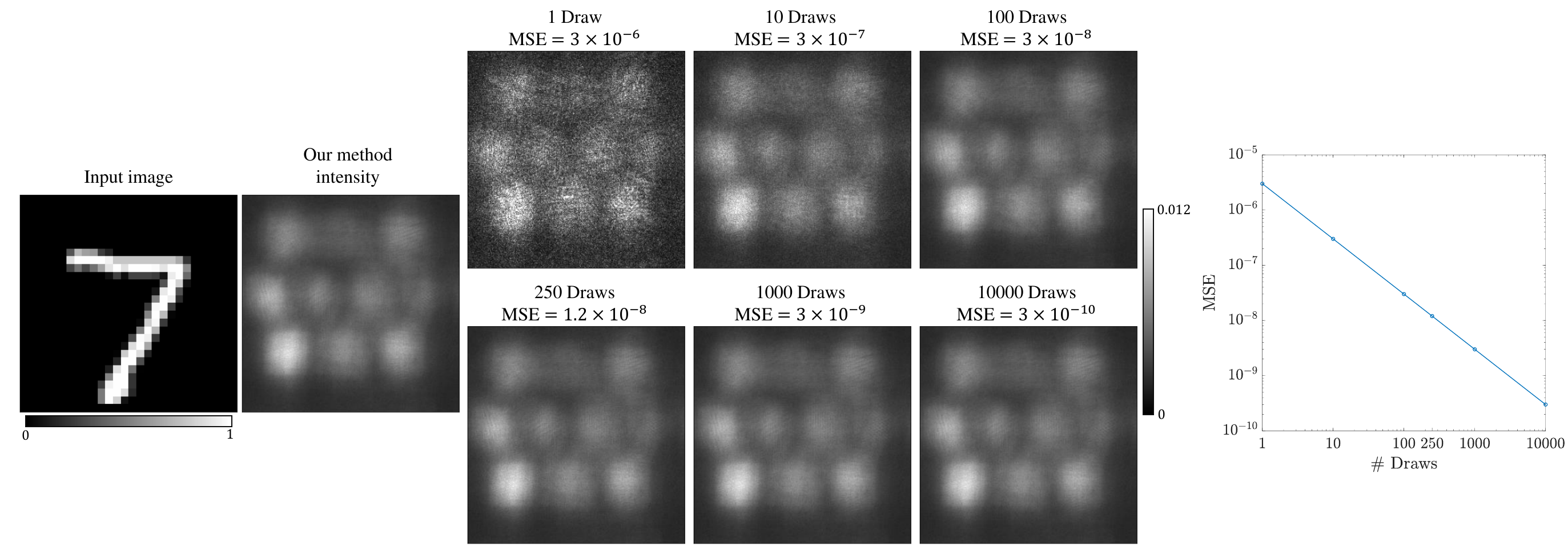}
   \caption{\textbf{Modal expansion comparison.} Comparison between the intensity results obtained by passing a given input image (the digit 7) through a diffractive network trained using our method with those obtained using modal expansion with different numbers of draws. We calculated the MSE between the intensity images at the output plane obtained using our method and those obtained using modal expansion with different numbers of draws. As the number of draws increases, the MSE decrease, as shown on the graph on the right. For a large number of draws, there is virtually no difference between the methods. In the displayed input image there are 116 pixels with non-zero values, meaning that in this example, our method is computationally equivalent to 116 draws.}
   \label{fig:draws_intens}
\end{figure}

Figure \ref{fig:draws_intens} shows a comparison between the intensity results obtained with the method of propagating separate pixels of the source, which we used in our experiments, and the modal expansion method. The plot shows the results for a varying number of draws, when passing a single example through a diffractive network trained with our method.  As $N_{d}$ grows larger, the squared difference between the intensity image obtained with our method and that obtained with modal expansion tends to zero.

We additionally compared the accuracy results of a diffractive network trained using our method with those obtained using modal expansion with $N_d=10000$. Not surprisingly, we obtain the same accuracy results with both methods. However, using modal expansion with $N_d=10000$ draws is significantly more time consuming.

We further trained diffractive networks using modal expansion with the same hyperparameters as our method. To fairly compare the two methods we set $N_d$ to the average number of non-zero pixels in the train set images, which is 149 (recall that in our method, we only propagate each nonzero pixel of the input image). Training diffractive networks with modal expansion led to better results for diffractive networks with a small number of diffractive layers. However, diffractive networks with a larger number of diffractive layers trained with our method outperformed those trained with modal expansion. For example, a diffractive network with 3 diffractive layers trained for object dimensions of $(W_o, H_o) = (280,280)$ $\mu$m with modal expansion achieved an accuracy of $87.6\%$ on the MNIST dataset while training with our method and the same hyperparameters, it achieved an accuracy of $88.9\%$. This interesting phenomenon, where modal expansion is less accurate in predicting the output intensity but not always worse for the purpose of training diffractive networks, seems to be associated with its inherent randomness. This is similar in nature to random perturbation methods like Dropout~\cite{srivastava2014dropout} and DropConnect~\cite{wan2013regularization}, which are known to significantly improve results when training digital neural networks. It is certainly possible that our method can also benefit from intentionally introducing some sort of randomness during training. However, we leave this avenue for future work.

\clearpage
\section{Nonlinearity due to field-intensity quadratic relation}
\label{sm:nonlinearity_partial}

In a linear optical system, the relation between the object's transmittance $t_o$ and the output intensity $I_\text{out}$ is always quadratic. Indeed, it is described by Eq.~(S25), which involves the quadratic term $t_o(x,y)t_{o}^{*}(x+\Delta x,y + \Delta y)$. However, under different spatial coherence levels, this quadratic relation takes on different forms. In particular, under spatially incoherent light, the output intensity becomes linear in the input intensity. In our simulations with the MNIST and FashionMNIST datasets, the object's transmittance function is nearly binary, so that the intensity at the object plane is roughly the same as the object's transmittance (up to a multiplicative constant). This implies that for spatially incoherent light, the network is nearly linear in its input. This may explain the deterioration of the network's performance as the light becomes less spatially coherent, as seen in Fig.~\ref{fig:partial}.

\begin{figure}[h]
  \centering
   \includegraphics[width=0.7\linewidth]{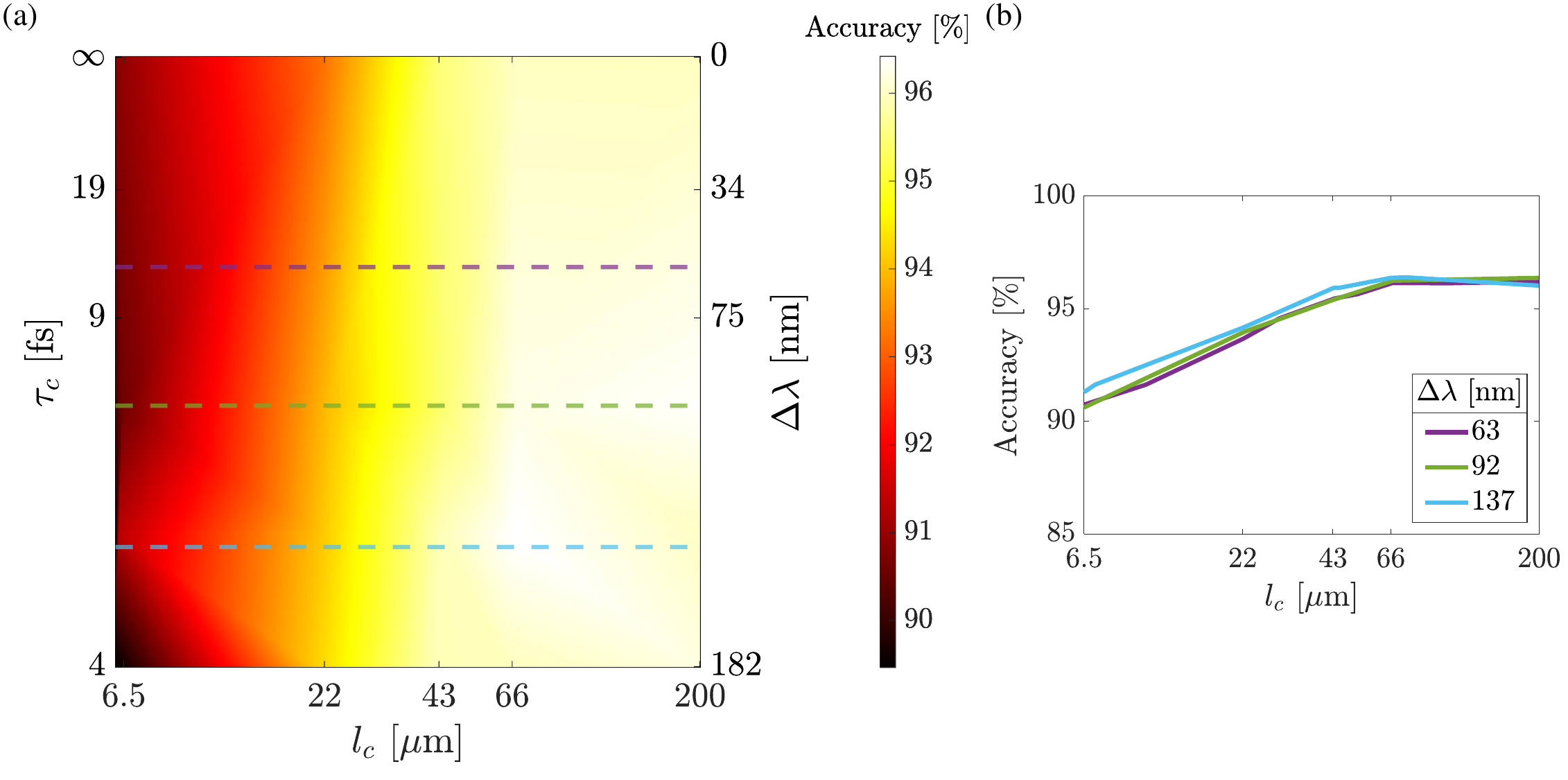}
   \caption{\textbf{Cross section of accuracy vs. spatial coherence.} (a) Taken from Fig.~\ref{fig:partial}a in the main text. Dashed colored lines indicate cross sections illustrated in (b). (b) Accuracy results vs. spatial coherence for the different cross sections of (a). The corresponding temporal coherence in wavelength bandwidth is given in the legend.}
   \label{fig:partial_nonlinear}
\end{figure}

The fact that the representation power of the network is weakest for spatially incoherent light, does not imply that its performance should necessarily improve monotonically as the light becomes more spatially coherent. Recall that Fig.~\ref{fig:partial}a shows how the performance varies with coherence, ranging from nearly incoherent to nearly coherent light. It can be seen that the performance generally improves as the spatial coherence grows larger, however it roughly levels off at around $66$ $\mu$m. To further examine this behavior, Fig.~\ref{fig:partial_nonlinear}b shows three cross-sections of Fig.~\ref{fig:partial_nonlinear}a, marked by colored dashed lines. Each color corresponds to a different temporal coherence condition, indicated by the wavelength bandwidth in the legend. Close inspection reveals that for broadband light (cyan curve), there is a subtle peak at approximately $66$ $\mu$m. Whether this minor advantage over spatially coherent light is larger under different network architectures is an interesting topic for future work.

In general, a quadratic nonlinearity is insufficient by itself for advanced computational tasks. Indeed, the strength of neural networks is associated with their universal approximation property~\cite{leshno1993multilayer}, which states that with a single hidden layer (a single nonlinearity, as in our case), they can approximate arbitrarily well any continuous function \textit{if and only if the activation function is non-polynomial}.

It is possible to improve the performance of optical neural networks by cascading multiple quadratic nonlinear layers, which leads to a high-order polynomial relation between the network's output and input. This was illustrated by Zhou et al.~\cite{zhou2021large} (see Fig.~S7 in their paper) using electro-optical layers.

\clearpage
\section{Simulating the forward propagation through the diffractive network}
\label{sm:forward_prop}

Propagation of a field through a linear diffractive network involves free-space propagation between consecutive diffractive layers and modulation of the optical field by each diffractive layer.

Free space propagation is simulated using the Rayleigh-Sommerfeld diffraction formulation and the angular spectrum method \cite{goodman2005introduction}. The Rayleigh-Sommerfeld transfer function is given by
\begin{equation} \label{eq:RS}  
H_{\text{R-S}}(f_{x},f_{y};z,\lambda) = \begin{cases}
    \exp\left\{j \frac{2\pi}{\lambda} z\sqrt{1-\lambda^2(f_x^2+f_y^2)}\right\} & \text{ if } \sqrt{f_x^2+f_y^2} \leq \frac{1}{\lambda},\\
    0 & \text{otherwise},
\end{cases}
\end{equation}
where $\lambda$ is the wavelength, $z$ is the propagation distance, $f_{x}$ and $f_{y}$ are the spatial frequencies along the $x$ and $y$ directions, respectively. Using this transfer function, we can write the electromagnetic after propagation by a distance $d$, as
\begin{equation}
    E(x,y;z+d,\lambda) = \mathcal{F}^{-1}\{\mathcal{F}\{E(x',y';z,\lambda)\} \cdot H_{\text{R-S}}(f_{x},f_{y};d,\lambda)\},
\end{equation}
where $\mathcal{F}$ and $\mathcal{F}^{-1}$ are the two-dimensional Fourier transform and inverse Fourier transform operations, respectively.

In our setting, each diffractive layer modifies only the phase of the field. It does so by multiplying the field by
\begin{equation} \label{eq:phase_modulation}
t_{\text{layer}}(x,y;\lambda) = \exp\{-j\phi(x,y;\lambda)\},   
\end{equation}
where $\phi(x,y;\lambda)$ is the learned phase modulation. The learned phase can be written as $\phi(x,y;\lambda) = \frac{2\pi n}{\lambda}h(x,y)$, where $h(x,y)$ is a height map and $n$ is the refractive index of the diffractive layer. Thus, the phase modulation for a specific wavelength $\lambda_{0}$ can be written as
\begin{equation} 
t_{\text{layer}}(x,y;\lambda_0)=\exp\left\{-j\frac{2\pi n}{\lambda_{0}}h(x,y)\right\}.
\end{equation}

Therefore, for a different wavelength, $\lambda_{i}\neq\lambda_{0}$, the phase modulation can be written as
\begin{equation} \label{eq:diff_lambda}
t_{\text{layer}}(x,y;\lambda_i)=\exp\left\{-j\frac{\lambda_{0}}{\lambda_{i}}\phi(x,y)\right\}. 
\end{equation}

At the end of the diffractive network, the detectors in the output plane measure the intensity, $I$, which is given by
\begin{equation}
I(x,y;z) = \sum_{\lambda}\left|E(x,y;z,\lambda)\right|^2.
\end{equation}

\clearpage

\section{Digital implementation and validation on simple settings}
\label{sm:digital}

As mentioned in \ref{sm:forward_prop}, we simulate the free space propagation using the angular spectrum method \cite{goodman2005introduction}, which involves the Fourier transform of the Rayleigh-Sommerfeld kernel, given in \eqref{eq:RS}. Since we perform digital calculations, this requires sampling Eq.~\eqref{eq:RS} on a dense enough grid. It is well known that to adequately sample the chirp function in Eq.~\eqref{eq:RS} the number of pixels in the sampled array, $N$, must satisfy
\begin{equation} \label{eq:sampling_criteria}
    N \ge \frac{\lambda z}{(\Delta x)^2},
\end{equation}
where $\lambda$ is the wavelength, $z$ is the propagation distance, and $\Delta x$ is the sampling interval in real space~\cite{li2007diffraction, voelz2011computational}.

In our implementation, we used a sampling interval $\Delta x$ of $10$ $\mu$m, a maximal wavelength $\lambda$ of $700$ nm, a propagation distance $z$ of $5$ cm, and $N=300$ pixels, which translates into a side length of $3$ mm. Substituting these values into Eq.~\eqref{eq:sampling_criteria} we get that the chirp function is adequately sampled. Specifically, we zero-padded the object image, whose original side length is $28$ pixels ($0.28$ mm) to $N=300$ pixels ($3$ mm) so as to avoid aliasing~\cite{voelz2011computational}.
\begin{figure}[h]
  \centering
   \includegraphics[width=\linewidth]{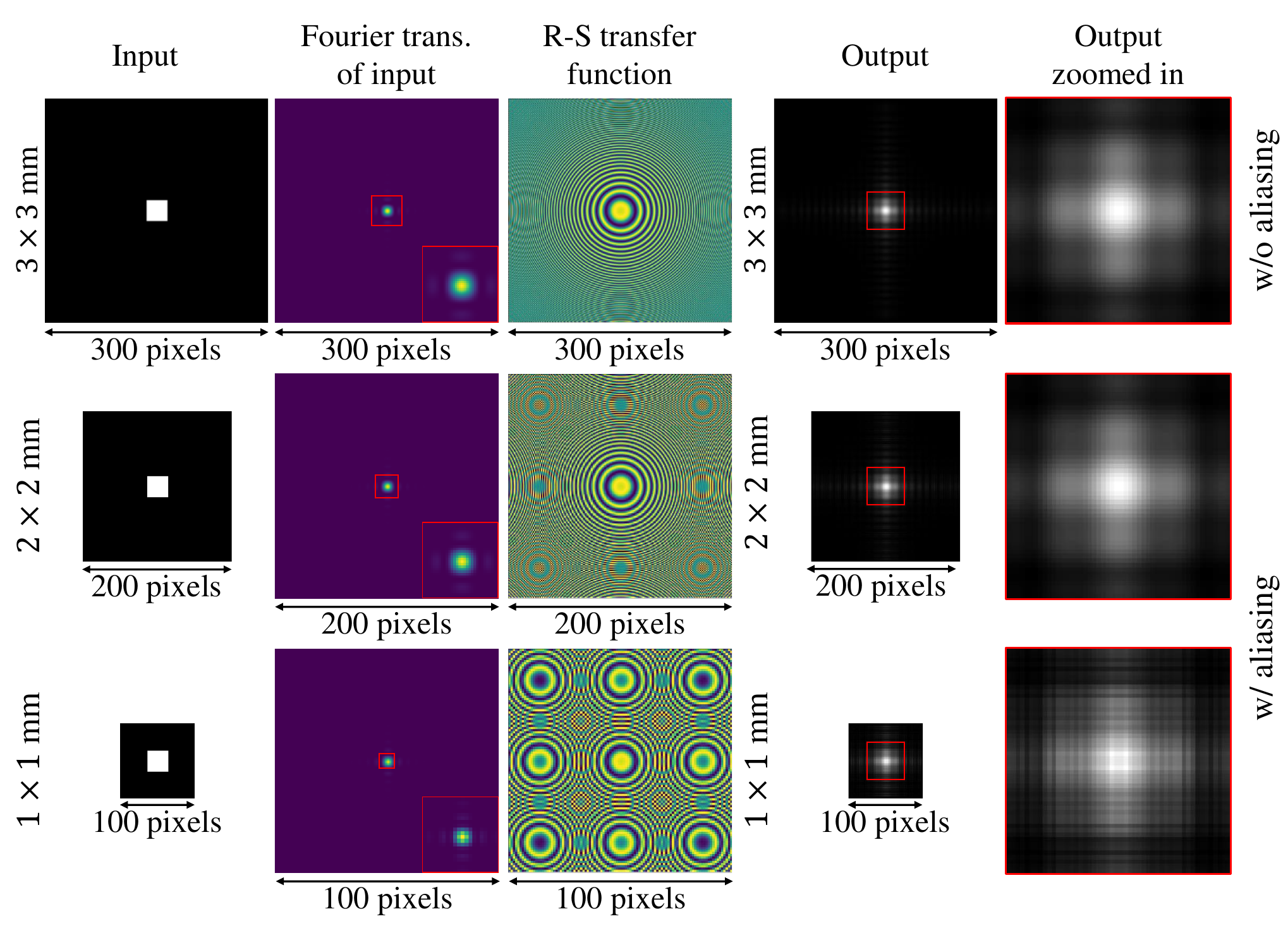}
   \caption{\textbf{Effect of padding and sampling on the propagation results of a square beam.} In each row the same square beam, with a side length of $0.28$ mm, is zero-padded to an array of $100$, $200$, or $300$ sizes, which translates into side sizes of $1$, $2$, or $3$ mm. The magnitude of the Fourier transform of the input, the real part of the Rayleigh-Sommerfeld transfer function and the resulting output intensity are illustrated for each padding scenario. A zoomed-in version of the magnitude of the Fourier transform of the input and the output are provided with red squares. The first row, where there is no aliasing on the Rayleigh-Sommerefeld transfer function, is the one we used in our numerical experiments. }
   \label{fig:sampling}
\end{figure}

Figure~\ref{fig:sampling} exemplifies the effect of padding and sampling on the propagation results of a square beam. In all rows, the input square beam has a side length of $28$ pixels ($0.28$ mm), the same as the entire object in our numerical experiments. We used a propagation distance of $z=5$ cm, as in our other numerical experiments, and took the wavelength to be $\lambda=550$ nm. The different rows depict the numerical results for different amounts of zero padding. Each square beam is propagated using the angular spectrum method described above. The third column shows the real part of the Rayleigh-Sommerfeld transfer function, following Eq.~\eqref{eq:RS}. 

It can be seen that insufficient padding leads to aliased sub-sampling of the the Rayleigh-Sommerfeld transfer function, which results in artifacts in the output intensity.  The first row in Fig.~\ref{fig:sampling}, in which there is no aliasing and no artifacts in the output intensity, is the setting we used.

\end{document}